\documentclass[reprint,aip,prl,amsmath,amssymb,reprint,twocolumn]{revtex4-2}

\usepackage{graphicx}                                   
\usepackage{dcolumn}                                    
\usepackage{bm}                                         
\usepackage{chemformula} 
\usepackage[utf8x]{inputenc} 
\usepackage[T1]{fontenc} 
\usepackage{mathrsfs}
\usepackage{amsmath}
\usepackage{braket}
\usepackage{amssymb}
\usepackage{appendix}
\usepackage{hyperref}
\usepackage{color}
\usepackage[normalem]{ulem}
\usepackage{dcolumn}
\usepackage{adjustbox}
\usepackage{tabularx}
\usepackage{tikz}
\usepackage{physics}
\usepackage{program}
\usepackage{graphicx}
\usepackage[normalem]{ulem}
\usetikzlibrary{arrows,shapes,positioning,shadows,trees}

\newcommand{\red}[1]{{\color{black} #1}}

\newcommand{\nl}{\nonumber \\}

\hypersetup{colorlinks=true, citecolor=blue, urlcolor=blue, linkcolor=blue}

\draft 

\begin{document}

\title{Toward Quantum-Aware Machine Learning: Improved Prediction of Quantum Dissipative Dynamics via Complex Valued Neural Networks} 
\author{Muhammad Atif}
\affiliation{School of Physics, Anhui University, Hefei, 230601, Anhui, China}  
\author{Arif Ullah}   
\email{arif@ahu.edu.cn}
\affiliation{School of Physics, Anhui University, Hefei, 230601, Anhui, China}  
\author{Ming Yang}  
\email{mingyang@ahu.edu.cn}
\affiliation{School of Physics, Anhui University, Hefei, 230601, Anhui, China}  

\date{\today}

\begin{abstract}
Accurately modeling quantum dissipative dynamics remains challenging due to environmental complexity and non-Markovian memory effects. Although machine learning provides a promising alternative to conventional simulation techniques, most existing models employ real-valued neural networks (RVNNs) that inherently mismatch the complex-valued nature of quantum mechanics. By decoupling the real and imaginary parts of the density matrix, RVNNs can obscure essential amplitude–phase correlations, compromising physical consistency. Here, we introduce complex-valued neural networks (CVNNs) as a physics-consistent framework for learning quantum dissipative dynamics. CVNNs operate directly on complex-valued inputs, preserve the algebraic structure of quantum states, and naturally encode quantum coherences. Through numerical benchmarks on the spin-boson model and few variants of the Fenna--Matthews--Olson complex, we demonstrate that CVNNs outperform RVNNs in convergence speed, training stability, and physical fidelity---including significantly improved trace conservation and Hermiticity. These advantages increase with system size and coherence complexity, establishing CVNNs as a robust, scalable, quantum-aware classical approach for simulating open quantum systems in the pre-fault-tolerant quantum era.
\end{abstract}
 
\maketitle
\section{Introduction}
Open quantum systems provide a fundamental framework for understanding how quantum states evolve under the influence of external environments (baths). Such system–environment interactions are central to numerous physical phenomena, including quantum information processing\cite{breuer2016colloquium}, quantum memory\cite{khodjasteh2013designing}, quantum transport\cite{cui2006quantum}, photosynthetic energy transfer\cite{zerah2021photosynthetic}, and biochemical processes such as proton tunneling in DNA\cite{slocombe2022open}. Conceptually, in the system-bath model, an open quantum system consists of a primary subsystem of interest coupled to an external bath. The dynamics of the subsystem are encoded in the reduced density matrix (RDM), which captures both intrinsic quantum evolution and the influence of environmental degrees of freedom.

Modeling environmental effects, however, is highly challenging due to the vast and often continuous degrees of freedom associated with the bath. Existing approaches span two broad categories: mixed quantum–classical treatments and fully quantum formulations. Mixed quantum–classical schemes\cite{Miller2001SCIVR, cotton2013symmetrical, liu2021unified, runeson2019spin, runeson2020generalized, mannouch2020partially, mannouch2020comparison, mannouch2022partially, tao2016multi, mannouch2023mapping, crespo2018recent, qiu2022multilayer} treat the system quantum mechanically while representing the environment through classical phase-space variables such as molecular vibrations or solvent motion, whose evolution follows Newtonian or Hamiltonian dynamics. This hybridization dramatically reduces computational cost by avoiding an explicit representation of the environment’s Hilbert space.

Fully quantum approaches, in contrast, place both the system and the environment within the quantum Hilbert space, enabling a complete treatment of coherence, dissipation, and entanglement. Path-integral-based methods such as the hierarchical equations of motion (HEOM)\cite{tanimura1989time, makarov1994path, su2023extended, yan2021efficient, gong2018quantum, xu2022taming, bai2024heom, wang2023simulating} and the quasiadiabatic propagator path integral (QUAPI)\cite{makarov1994path, makri2023quantum} rigorously capture non-Markovian memory effects. Similarly, quantum master equation approaches--including the stochastic equation of motion (SEOM)\cite{han2019stochastic, han2020stochastic, ullah2020stochastic, chen2022simulation, dan2022generalized, stockburger2016exact} and the generalized master equation\cite{lyu2023tensor, liu2018exact}---provide alternative fully quantum frameworks for describing environmental interactions.

Despite their success, both categories face well-known limitations. Mixed quantum–classical approaches may break detailed balance\cite{Schmidt2008equilibrium, ellipsoid, thermalization} or fail to capture subtle quantum correlations\cite{Mannouch2024coherence}. Fully quantum methods, while theoretically rigorous, often become prohibitively expensive for strong coupling regimes, long memory times, or large environmental modes, where fine temporal resolution is required to ensure numerical stability.

In response to these challenges, machine learning (ML) has recently emerged as a promising alternative for modeling complex spatiotemporal dynamics in quantum systems.\cite{ullah2025machine, ullah2021speeding, ullah2022predicting, ullah2022one, rodriguez2022comparative, herrera2021convolutional, ge2023four, zhang2023excited, wu2021forecasting, lin2022automatic, bandyopadhyay2018applications, yang2020applications,lin2022trajectory,tang2022fewest,  shakiba2024ml_spin_relaxation, lin2024enhancing, zeng2024sophisticated, long2024quantum, cao2024neural, zhang2024nonmarkov, zhang2024ai, herrera2024short, ullah2024mlqd, PhysRevResearch.7.L012013, ullah2025short, zhang2025neural, zhao2025multi} Existing work, which rely exclusively on real-valued neural networks (RVNNs), have demonstrated strong predictive capabilities in forecasting quantum dynamics from the past history of the system and in learning direct mappings between system parameters, time, and the corresponding quantum states.

Despite these successes, a fundamental representational tension exists: quantum mechanics is intrinsically formulated within a complex-valued Hilbert space. This complex structure is not merely a mathematical convenience but encodes essential physical information, particularly in the phase relationships that define quantum coherence and interference. A parallel and compelling development across diverse scientific and engineering domains---from advanced signal processing and computational imaging to materials discovery and even emerging architectures for large language models\cite{renou2021quantum, chen2022ruling, li2022testing, wang2025ifairy, cole2021analysis, igelnik1999quantum, trabelsi2018deep, bassey2021survey, zhang2025complex, dong2024quantum, barrachina2021complex, yang2017finance}---underscores that complex-valued neural networks (CVNNs) possess a unique, intrinsic capacity to capture phase-dependent structures and continuous rotational symmetries. These are features that RVNNs, by their architectural design, inherently struggle to represent. The standard practice in RVNNs is to decouple the real and imaginary components of complex-valued data, processing them through separate, parallel real-valued pathways. This bifurcation risks discarding or obscuring the critical correlations that are holistically encoded within the complex geometry of quantum states, potentially limiting physical fidelity and generalization.

While quantum neural networks would, in principle, offer the most natural framework for quantum dynamics learning, current noisy intermediate-scale quantum (NISQ)-era devices remain too limited in scale and fidelity to outperform classical approaches. This motivates the search for classical architectures that can emulate essential quantum features without requiring quantum hardware. Given that complex numbers serve as a compact and faithful representation of quantum structure,\cite{renou2021quantum, chen2022ruling, li2022testing} CVNNs emerge as a promising intermediary between RVNNs and fully quantum models.

In this work, we conduct a systematic comparison between CVNNs and RVNNs for learning and forecasting quantum dissipative dynamics. Using the spin-boson (SB) model and several variants of the Fenna--Matthews--Olson (FMO) complex, we show that CVNNs converge faster, exhibit greater training stability, and yield predictions with improved physical fidelity---including superior trace conservation and Hermiticity---relative to RVNNs. Importantly, these advantages become increasingly pronounced with growing system size and coherence complexity, highlighting the scalability of the CVNN framework. Together, our results establish CVNNs as a powerful classical surrogate for quantum-aware learning and a practical pathway for modeling open quantum systems in the pre-fault-tolerant quantum era.

\section{Theory and Methodology}

We consider an open quantum system $\mathrm{S}$ with Hilbert space $\mathcal{H}_{\mathrm{S}}$ of dimension $n$, interacting with an environment $\mathrm{E}$ with Hilbert space $\mathcal{H}_{\mathrm{E}}$ of dimension $d$. The joint system $\mathrm{S} + \mathrm{E}$ evolves unitarily under the Liouville–von Neumann equation $(\hbar = 1)$
\begin{equation}
    \dot{\rho}(t) = -\mathrm{i}[H, \rho(t)],
\end{equation}
where $H$ is the total Hamiltonian acting on the tensor product space $\mathcal{H}_{\mathrm{S}}\otimes\mathcal{H}_{\mathrm{E}}$, and $\rho(t)\in\mathbb{C}^{(n d)\times(nd)}$ denotes the full density matrix of system plus environment. Given an initially factorized state
\begin{equation}
    \rho(0) = \rho_{\mathrm{S}}(0)\otimes\rho_{\mathrm{E}}(0),
\end{equation}
the unitary propagator $U(t)=e^{-iHt}\in\mathbb{C}^{(nd)\times(nd)}$ generates the reduced dynamics of the subsystem through the partial trace,
\begin{equation}
    \rho_{\mathrm{S}}(t)
= \operatorname{Tr}_{\mathrm{E}}\big[ U(t)\rho(0)U^\dagger(t)\big],
\end{equation}
where $\rho_{\mathrm{S}}(t)\in\mathbb{C}^{n\times n}$ represents the RDM of the system. The map $\rho_{\mathrm{S}}(0)\mapsto \rho_{\mathrm{S}}(t)$ is typically non-Markovian and cannot, in general, be expressed as a closed local evolution equation. Instead, one may write a generalized master equation of the form
\begin{equation}
    \dot{\rho}_{\mathrm{S}}(t)
= -\mathrm{i}[H_{\mathrm{S}}, \rho_{\mathrm{S}}(t)]
+ \mathcal{R}\left[{\rho_{\mathrm{S}}(s)}_{0\le s\le t}\right],
\end{equation}
where $H_{\mathrm{S}}$ acts on the system alone, and $\mathcal{R}$ is a superoperator encoding dissipation, decoherence, and memory. Abstractly, the reduced dynamics can therefore be represented as a history-dependent quantum map
\begin{equation}
    \rho_{\mathrm{S}}(t)
= \Phi_t\left[\{\rho_{\mathrm{S}}(s)\}_{0\le s < t} \right],
\end{equation}
where $ \Phi_t$ is, in principle, a completely positive and trace-preserving (CPTP) operator-valued functional on the space of continuous matrix-valued trajectories.

From ML perspective, the goal is to learn and approximate the non-Markovian evolution operator $\Phi_t$ using a parametrized model. To formalize this, let ${\rho_{\mathrm{S}}(t_{k-k'}),\dots,\rho_{\mathrm{S}}(t_k)}\subset \mathbb{C}^{n\times n}$ denote a discrete-time history window of length $k' +1$, containing the most recent RDMs. We define a recursive operator
\begin{equation}
    \mathcal{C}_{\mathrm{rec}} : (\mathbb{C}^{n\times n})^{k'+1} \longrightarrow (\mathbb{C}^{n\times n})^N
\end{equation}
which maps the sequence of past RDMs to the next $N$ future RDMs,
\begin{equation}
    \mathcal{C}_{\mathrm{rec}}\big(\rho_{\mathrm{S}}(t_{k-k'}),\dots,\rho_{\mathrm{S}}(t_k)\big) = \big(\rho_{\mathrm{S}}(t_{k+1}),\dots,\rho_{\mathrm{S}}(t_{k+N})\big).
\end{equation}

Formally, if we denote the discrete trajectory at step $k$ by the vector-valued sequence
\begin{equation}
    \mathbf{R}_k = \big(\rho_{\mathrm{S}}(t_{k-k'}), \dots, \rho_{\mathrm{S}}(t_k)\big) \in (\mathbb{C}^{n\times n})^{k'+1},
\end{equation}
then the recursion reads
\begin{align}
    \mathbf{R}_{k+N} & = \big(\rho_{\mathrm{S}}(t_{k-k'+N+1}),\dots,\rho_{\mathrm{S}}(t_{k+N})\big) \nl 
    & = \mathrm{shift}_N\big(\mathbf{R}_k, \mathcal{C}_{\mathrm{rec}}(\mathbf{R}_k)\big),
\end{align}
where $\mathrm{shift}_N$ discards the oldest $N$ matrices and appends the newly predicted ones. Iterating this mapping produces the full predicted trajectory of the reduced dynamics
$\rho_{\mathrm{S}}(t_0), \rho_{\mathrm{S}}(t_1), \dots, \rho_{\mathrm{S}}(t_T)$ providing a recursive ML-based approximation of the true non-Markovian propagator $\Phi_t$ over the time interval of interest.

\subsection{Real and complex valued neural networks}

To implement the recursive operator $\mathcal{C}_{\mathrm{rec}}$ we employ both RVNNs and CVNNs. Let $\rho_{\mathrm{S}}(t)\in\mathbb{C}^{n\times n}$ denote the RDM of the system. Because $\rho_{\mathrm{S}}$ is Hermitian $(\rho_{kl}=\rho_{lk}^*)$, all independent information resides in the upper-triangular part including the diagonal; we denote this vectorized object by $\rho_{\mathrm{S}}^{\rm up}\in\mathbb{C}^{n(n+1)/2}$. Using the upper-triangular representation reduces input dimensionality and enforces Hermiticity by construction (the lower triangle is recovered by conjugation).

In the RVNN formulation, all inputs, weights, and activations lie in $\mathbb{R}$, requiring a real decomposition of the complex-valued upper-triangular data. For each off-diagonal element $\rho_{kl}=a_{kl}+ib_{kl}$ with $k<l$, the real and imaginary components $a_{kl}, b_{kl}$ are treated as separate degrees of freedom, while diagonal entries—being real and having no imaginary component—contribute only their real values. Flattening this representation yields a real vector $\mathbf{x}^{\mathrm{up}}\in\mathbb{R}^{n^2}$. An RVNN layer then performs the mapping
\begin{equation}
    \mathbf{y}=\sigma\big(\mathbf{W}\mathbf{x}^{\rm up}+\mathbf{b}\big),\qquad
\mathbf{W}\in\mathbb{R}^{m\times n^2},\ \mathbf{b}\in\mathbb{R}^m,
\end{equation}
with $\sigma$ a real-valued nonlinearity (ReLU, $\tanh$, etc.). This real embedding treats $\Re$ and $\Im$ components as independent channels and therefore lacks the natural complex geometry: real multiplication effects independent scaling of each channel but cannot implement planar rotations or amplitude–phase coupling in a single algebraic operation. Consequently RVNNs are not $\mathbb{C}$-linear and do not preserve phase structure; they learn dynamics in a surrogate real vector space rather than the native complex Hilbert space, which can limit expressivity for long-time coherent phenomena.

CVNNs, in contrast, operate directly on complex-valued inputs and preserve the algebraic structure of complex state evolution. We encode the upper-triangular RDM directly as a complex vector $\mathbf{z}=\rho_{\mathrm{S}}^{\rm up}\in\mathbb{C}^{n(n+1)/2}$. For diagonal entries, which are purely real, we embed them uniformly into $\mathbb{C}$ by mapping $a\mapsto a+ia$ at the input; at the output, this additive phase is discarded to maintain Hermiticity. A CVNN layer computes 
\begin{equation}
    \mathbf{f}_{\mathbb{C}}(\mathbf{z})=\sigma_{\mathbb{C}}(\mathbf{W}\mathbf{z}+\mathbf{b}),\qquad
\mathbf{W}\in\mathbb{C}^{m\times n(n+1)/2},\ \mathbf{b}\in\mathbb{C}^m,
\end{equation}
where $\sigma_{\mathbb{C}}$ is a complex-valued activation applied component-wise. Writing $\mathbf{z}=\mathbf{u}+i\mathbf{v}$, $\mathbf{W}=\mathbf{W}_r+i\mathbf{W}_i$ and $\mathbf{b}=\mathbf{b}_r+i\mathbf{b}_i$, the linear part is represented in real block form as
\begin{equation}
    \begin{bmatrix}
\Re(\mathbf{Wz}+\mathbf{b}) \\
\Im(\mathbf{Wz}+\mathbf{b})
\end{bmatrix}
=
\begin{bmatrix}
\mathbf{W}_r & -\mathbf{W}_i \\
\mathbf{W}_i & \mathbf{W}_r
\end{bmatrix}
\begin{bmatrix}
\mathbf{u}\\ \mathbf{v}
\end{bmatrix}
+
\begin{bmatrix}
\mathbf{b}_r\\ \mathbf{b}_i
\end{bmatrix}.
\end{equation}

This structured block matrix represents true complex multiplication, where the real and imaginary parts transform jointly rather than independently, preserving amplitude–phase coupling before the activation is applied.

Geometrically, each component of $\rho_{\mathrm{S}}^{\rm up}$, say $\rho_{kl}=u_{kl}+iv_{kl}$, corresponds to a two-dimensional vector $(u_{kl},v_{kl})$ in the complex plane. Multiplication by a complex weight $W_{kl}^c=W_{kl}^r+i W_{kl}^i$ performs a rotation and a scaling in this plane: the angle of rotation is
\begin{equation}
    \arg(W_{kl}^c) = \arctan\big(W_{kl}^i / W_{kl}^r\big),
\end{equation}
and the magnitude,
\begin{equation}
    \left\lvert W_{kl}^c  \right \lvert=\sqrt{(W_{kl}^r)^2+(W_{kl}^i)^2},
\end{equation}
determines the dilation (scaling). Adding a complex bias translates the vector in the same plane. Thus, each CVNN neuron implements a transformation that simultaneously rotates, scales, and translates the complex-valued entries of the RDM—precisely the class of planar transformations compatible with complex linearity. No RVNN can generate such rotations in a single step, because real weights act independently on $\Re$ and $\Im$ channels and therefore cannot encode angle-preserving transformations. This geometric capacity to preserve and manipulate phase relationships is precisely why CVNNs are more faithful to reproduce phase-sensitive, interference-driven, and non-Markovian quantum dynamics.

\subsection{Weight initialization in real-valued and complex-valued neural layers}

Accurate modeling of quantum dynamical maps such as the recursive operator $\mathcal{C}_{\mathrm{rec}}$ requires careful control over signal propagation in both real-valued and complex-valued neural architectures. Because RVNNs operate in $\mathbb{R}^d$ whereas CVNNs operate in $\mathbb{C}^d$, their initialization schemes must be adapted to the geometry and variance structure of these two spaces. The real and imaginary channels in a CVNN jointly determine the magnitude and phase of the signal, so the initialization must preserve neutrality in the complex plane while matching the variance of the underlying real-valued baseline. 

In RVNN, we initialize each weight using the He (Kaiming) uniform distribution\cite{he2015delving} with variance $\mathrm{Var}[W_{kl}] = 2/n_{\mathrm{in}}$ with $n_{\mathrm{in}}$ as a number of input features. We compute the standard deviation $\sigma=\sqrt{2/n_{\mathrm{in}}}$ and sample from a symmetric uniform distribution $[-a, a]$ where $a = \sqrt{3}\sigma$. Since a uniform distribution on $[-a,a]$ has variance $a^2/3$, this ensures the outgoing variance matches the desired He level. Because the recursive operator repeatedly applies real-valued linear maps followed by nonlinearities, this initialization keeps the forward signal variance approximately stable and prevents vanishing gradients in early training. All real-valued biases are initialized to zero, a standard practice that allows the nonlinearities to determine the initial operating point.

The complex-valued layers in a CVNN require a different treatment because a complex weight $W_{kl} = W_{kl}^r + i W_{kl}^i$ is effectively a 2D vector whose variance decomposes across two components. If both real and imaginary parts were initialized with the real He variance $2/n_{\mathrm{in}}$, the effective complex variance $\mathbb{E}\left\lvert W_{kl} \right\lvert^2 = \mathrm{Var}[W_{kl}^r] + \mathrm{Var}[W_{kl}^i]$ would double, leading to unstable magnitude growth.\cite{trabelsi2018deep} To avoid this, we assigns each component the reduced variance $\mathrm{Var}[W_{kl}^r] = \mathrm{Var}[W_{kl}^i] = 1/n_{\mathrm{in}}$. The resulting complex variance is then
\begin{equation}
    \mathbb{E}\left\lvert W_{kl} \right\lvert^2 = \frac{1}{n_{\mathrm{in}}} + \frac{1}{n_{\mathrm{in}}}
= \frac{2}{n_{\mathrm{in}}},
\end{equation}
matching the real He initialization only after combining the two channels. We achieve this by setting $\sigma = \sqrt{1/n_{\mathrm{in}}}$ and sampling $W_{kl}^r$ and $W_{kl}^i$ independently from a uniform distribution on $[-a,a]$ with $a = \sqrt{3}\sigma$. This construction preserves phase neutrality: the distribution of $W_{kl}$ has no preferred direction in the complex plane, ensuring no initial bias in rotational transformations. Such neutrality is critical for maintaining phase coherence, because the linear operator induced by a complex weight, $z \mapsto W_{kl} \, z = (W_{kl}^r + iW_{kl}^i)(u_{kl}+iv_{kl})$, corresponds to a rotation by $\arg(W_{kl})$ and a scaling by $\left\lvert W_{kl} \right\lvert$. Any imbalance in the initial variances of $W_{kl}^r$ and $W_{kl}^i$ would distort this geometry, bias the angular distribution, or amplify certain directions in the complex plane. By splitting the variance equally between the two channels, the magnitude distribution remains normalized and the initial transformations act as unbiased complex scalings and rotations. Coming to Biases, in CVNN implementation, we set both the real and imaginary parts to zero, ensuring that the early-stage operator acts without translation, consistent with the affine-free structure of quantum channels.
%
%
\begin{figure*}
\includegraphics[width=\linewidth]{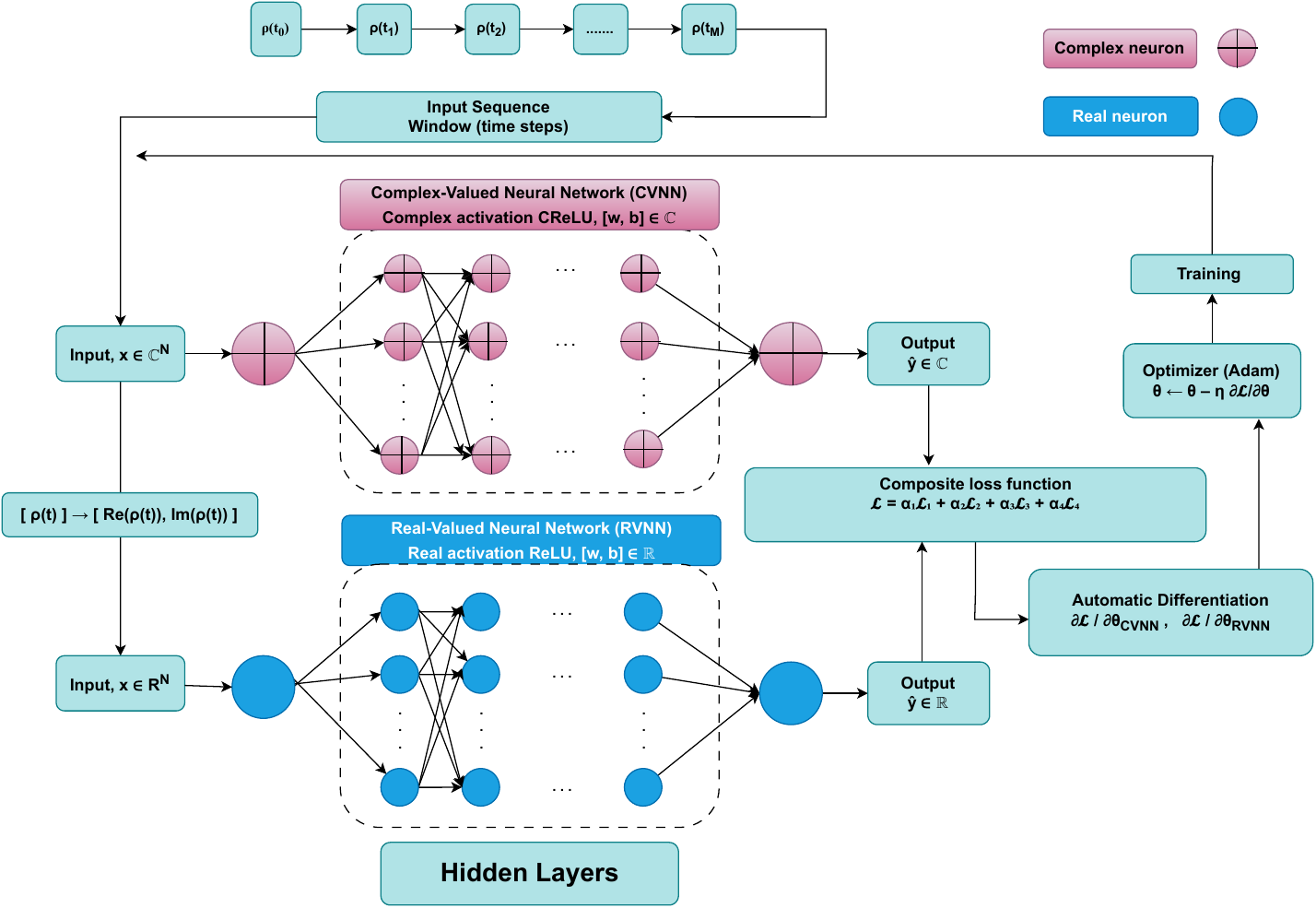}
\caption{A schematic flowchart illustrating the general training process for both CVNN and RVNN models. }
\label{fig:flowchart}
\end{figure*}
%

\section{Systems of Interest}

To evaluate the efficiency and quantum-awareness of RVNNs and CVNNs, we benchmark both architectures on representative open quantum systems that capture essential features of coherent evolution, dissipation, and system–bath interactions. Our first testbed is the paradigmatic SB model, which describes a two-level quantum system (qubit) interacting with a bosonic environment composed of infinitely many harmonic modes. The composite Hamiltonian is
\begin{equation}
    \mathbf{H} = \varepsilon \sigma_{z} + \Delta \sigma_{x}
 + \sum_{k} \omega_{k} b_{k}^{\dagger} b_{k}
+  \sigma_{z} \sum_{k} c_{k}(b_{k}^{\dagger} + b_{k}),
\end{equation}
  where $\varepsilon$ denotes the energy splitting of the two-level system, $\Delta$ is the tunneling amplitude between states, and $\sigma_{z},\sigma_{x}$ are Pauli matrices. The bosonic bath is defined through harmonic-mode operators $b_{k}, b_{k}^{\dagger}$ with mode frequencies $\omega_{k}$, coupled to the system with strength $c_{k}$. Environmental influence is encoded via the Debye spectral density,
\begin{equation}
      J(\omega)= 2\lambda \frac{\gamma \omega}{\omega^{2}+\gamma^{2}},
\end{equation}
  which introduces a reorganization energy $\lambda$ and a characteristic environmental relaxation timescale $\tau = 1/\gamma$. 

As a complementary and more structurally complex system, we also consider the FMO complex, a pigment–protein complex central to energy transfer in green sulfur bacteria. Each monomer of the FMO trimer contains seven or eight bacteriochlorophyll sites, forming a network through which electronic excitations propagate with remarkably high efficiency. The dynamics are modeled using the Frenkel exciton Hamiltonian,
\begin{equation}
    \begin{aligned}
\mathbf{H} =&
\sum_{n=1}^{N} \ket{n}\varepsilon_n \bra{n}
+ \sum_{n\neq m}\ket{n} J_{nm} \bra{m} \\
&+ \sum_{n,k} \left( \tfrac12 P_{k,n}^2 + \tfrac12 \omega_{k,n}^2 Q_{k,n}^2 \right) I \\
& - \sum_{n,k} \ket{n} c_{k,n} Q_{k,n} \bra{n}
+ \sum_{n} \ket{n}\lambda_n \bra{n}.
  \end{aligned}
\end{equation}

  Here, $\varepsilon_n$ denotes the site energy, $J_{nm}$ the electronic coupling between sites, and $Q_{k,n}$, $P_{k,n}$, and $\omega_{k,n}$ describe the coordinate, momentum, and frequency of environmental modes coupled to site $n$ with strength $c_{k,n}$. The reorganization energy $\lambda_n$ accounts for environmentally induced shifts in site energies and $I$ is the identity matrix for dimensional compatibility. For consistency and physical relevance, the environmental spectral density is again taken to be Debye with identical parameters across all sites.

\section{Results and Discussion}

In this work, we benchmark the performance of CVNNs against their real-valued counterparts (RVNNs) across four representative quantum dynamical systems of increasing Hilbert-space dimension: (i) the SB model, (ii) a hypothetical 4-site FMO model describing the population dynamics of the first four bacteriochlorophyll (BChl) sites of a 7-site FMO complex, (iii) the full 7-site FMO complex, and (iv) an 8-site extension of the FMO system. This systematic variation in system size allows us to assess the scalability and representational advantages of CVNNs relative to RVNNs.

For each system, we independently trained two neural network models—a RVNN and a CVNN—using identical training data. Both models adopt a fully connected sequence-to-sequence architecture consisting of five layers: one input layer, three hidden layers, and one output layer. In the RVNN, all layers employ real-valued weights and biases together with the ReLU activation function. In contrast, the CVNN is constructed using custom complex-valued linear layers with complex weights and biases, \red{combined with a complex ReLU (CReLU) activation function that applies the ReLU operation separately to the real and imaginary components}. \red{To ensure a fair comparison, we capped the number of trainable parameters in the CVNN at a level approximately equal to or below that of the RVNN.  Detailed architectures are listed in Tables~S1 and S2 (Section~S2 of the Supporting Information)}. Moreover, both network architectures are implemented from scratch within the PyTorch framework, allowing full control over the operations. A schematic overview of the training and optimization workflow for both models is shown in Fig.~\ref{fig:flowchart}.

To guarantee that the predicted RDMs satisfy the fundamental physical constraints of quantum mechanics—namely trace conservation, positive semi-definiteness, and bounded eigenvalues—we employ a composite loss function of the form
\begin{align}
\label{eq:8}
\mathcal{L}=\alpha_{1}\mathcal{L}_{1}+\alpha_{2}\mathcal{L}_{2}+\alpha_{3}\mathcal{L}_{3}+\alpha_{4}\mathcal{L}_{4},
\end{align}
where each term enforces a specific physical or numerical requirement.

The primary accuracy term, $\mathcal{L}_{1}$, is defined as the mean squared error (MSE) between the predicted RDM elements $\rho_{\rm S}(t)$ and the corresponding reference values $\tilde{\rho}_{\rm S}(t)$,
\begin{align}
\label{eq:9}
\mathcal{L}_{1}=\frac{1}{N_{t}\cdot n^{2}}\sum_{t=1}^{N_{t}}\sum_{i,j=1}^{n}
\left(\tilde{\rho}_{\rm S,ij}(t)-\rho_{\rm S,ij}(t)\right)^{2},
\end{align}
where $N_{t}$ denotes the total number of time steps and $n$ is the dimension of the system Hilbert space.

Trace preservation of the density matrix is enforced through the penalty term
\begin{align}
\label{eq:10}
\mathcal{L}_{2}=\frac{1}{N_{t}}\sum_{t=1}^{N_{t}}\left(\mathrm{Tr}\rho_{S}(t)-1\right)^{2},
\end{align}
which penalizes deviations of the predicted trace from unity at each time step.

The positive semi-definiteness of the RDM is imposed by penalizing negative eigenvalues $\mu_i(t)$ via
\begin{align}
\label{eq:11}
\mathcal{L}_{3}=\frac{1}{N_{t}\cdot n}\sum_{t=1}^{N_{t}}\sum_{i=1}^{n}
\max\left(0,-\mu_{i}(t)\right)^{2}.
\end{align}
In addition, physically admissible density matrices require all eigenvalues to lie within the interval $[0,1]$. This constraint is enforced through
\begin{align}
\label{eq:12}
\mathcal{L}_{4}=\frac{1}{N_{t}\cdot n}\sum_{t=1}^{N_{t}}\sum_{i=1}^{n}
\left(\mathrm{clip}(\mu_{i}(t),0,1)-\mu_{i}(t)\right)^{2},
\end{align}
where the clipping operation is defined as
\begin{align}
\label{eq:13}
\mathrm{clip}(\mu_{i}(t),0,1)=
\begin{cases}
0, & \mu_{i}(t)<0, \\
\mu_{i}(t), & 0 \le \mu_{i}(t) \le 1, \\
1, & \mu_{i}(t)>1.
\end{cases}
\end{align}

The relative importance of each constraint is controlled by the weighting coefficients $\alpha_{1}$--$\alpha_{4}$. In all simulations reported here, we use $\alpha_{1}=\alpha_{2}=1.0$, $\alpha_{3}=2.0$, and $\alpha_{4}=3.0$. Collectively, the loss terms $\mathcal{L}_{1}$–$\mathcal{L}_{4}$ ensure accurate reproduction of the reference dynamics while rigorously enforcing the fundamental physical properties of RDMs.

To train and evaluate the proposed neural network models, we employ high-quality reference data generated using numerically exact or well-established open quantum system approaches. For the SB model, training data are obtained from the openly accessible QD3SET-1 database\cite{ullah2023qd3set}. The dataset spans a four-dimensional parameter space, denoted by $\mathcal{D}_{\mathrm{sb}}$, defined by the normalized system–bath parameters $(\varepsilon/\Delta, \lambda/\Delta, \gamma/\Delta, \beta\Delta)$, which respectively characterize the bias-to-tunneling ratio, bath reorganization energy, bath relaxation rate, and inverse temperature. In total, $\mathcal{D}_{\mathrm{sb}}$ comprises 1000 independent simulations, with 500 trajectories for the symmetric case and 500 for the asymmetric case. For each parameter combination, the RDM dynamics are computed using the HEOM formalism\cite{tanimura1989time,shi2009efficient,xu2022taming,chen2022universal}, ensuring numerically accurate reference trajectories.

For the 7-site and 8-site FMO complexes, training data are likewise sourced from the QD3SET-1 database. These datasets describe excitation energy transfer dynamics initiated at site~1 and site~6 for the 7-site complex, with site~8 additionally included for the 8-site system. The associated parameter space $(\lambda, \gamma, T)$ spans bath reorganization energy, relaxation rate, and temperature, yielding a total of 500 simulations for each initial excitation scenario. The quantum dynamics are propagated using a trace-conserving local thermalizing Lindblad master equation (LTLME)\cite{mohseni2008environment}. The system Hamiltonians are parameterized following the models of Adolphs and Renger for the 7-site FMO complex\cite{adolphs2006proteins} and Jia \emph{et al.} for the 8-site extension\cite{jia2015hybrid}. \red{A detailed discussion of the parameter ranges and sampling scheme is provided in Section~S2 of the Supporting Information.}

Since a 4-site FMO model is not available in the QD3SET-1 database, we generate the corresponding dataset independently using the same LTLME framework and bath parameters as employed for the 7-site FMO complex, with the dynamics initialized by an excitation localized on site~1. This construction ensures consistency across all FMO prototypes considered in this study.

To improve training efficiency and ensure representative coverage of the parameter space, farthest point sampling\cite{dral2019mlatom,ullah2022predicting} is employed to select a subset of trajectories from each dataset \red{(See Section~S2 of the Supporting Information)}. For the SB model, we consider only the symmetric case, comprising 500 trajectories from $\mathcal{D}_{\mathrm{sb}}$. For each of the four systems---the SB model and the three FMO prototypes---we then select 400 trajectories for training via farthest point sampling, with the remaining trajectories reserved for out-of-sample testing.

\begin{figure}
\includegraphics[width=0.5\textwidth]{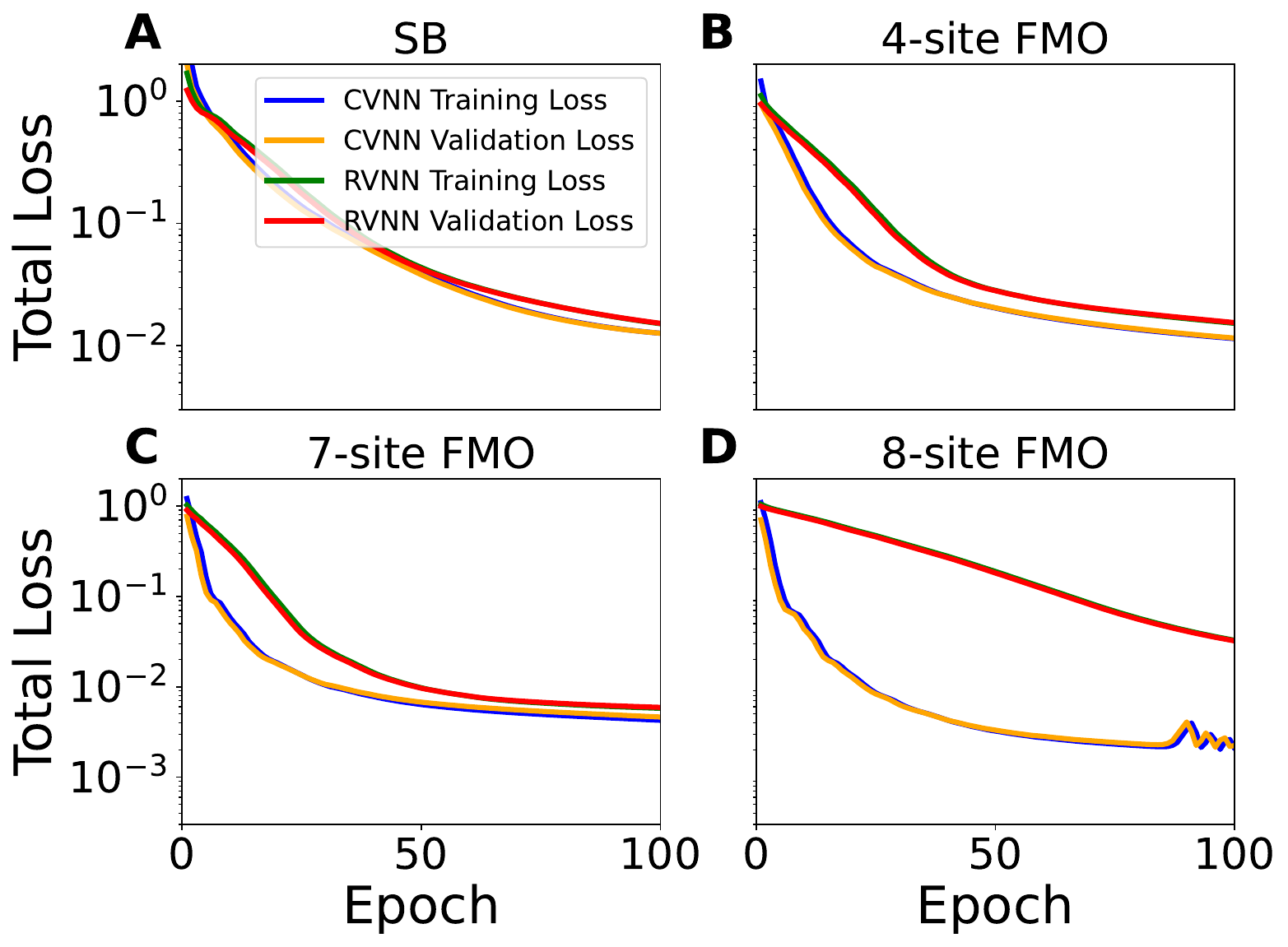}
\caption{\red{A comparison of training and validation loss curves versus epoch number for the CVNN and RVNN models. Results are shown for the SB model and all three prototypes of the FMO complex.}}
\label{fig:loss}
\end{figure}

\begin{figure}
\includegraphics[width=0.45\textwidth]{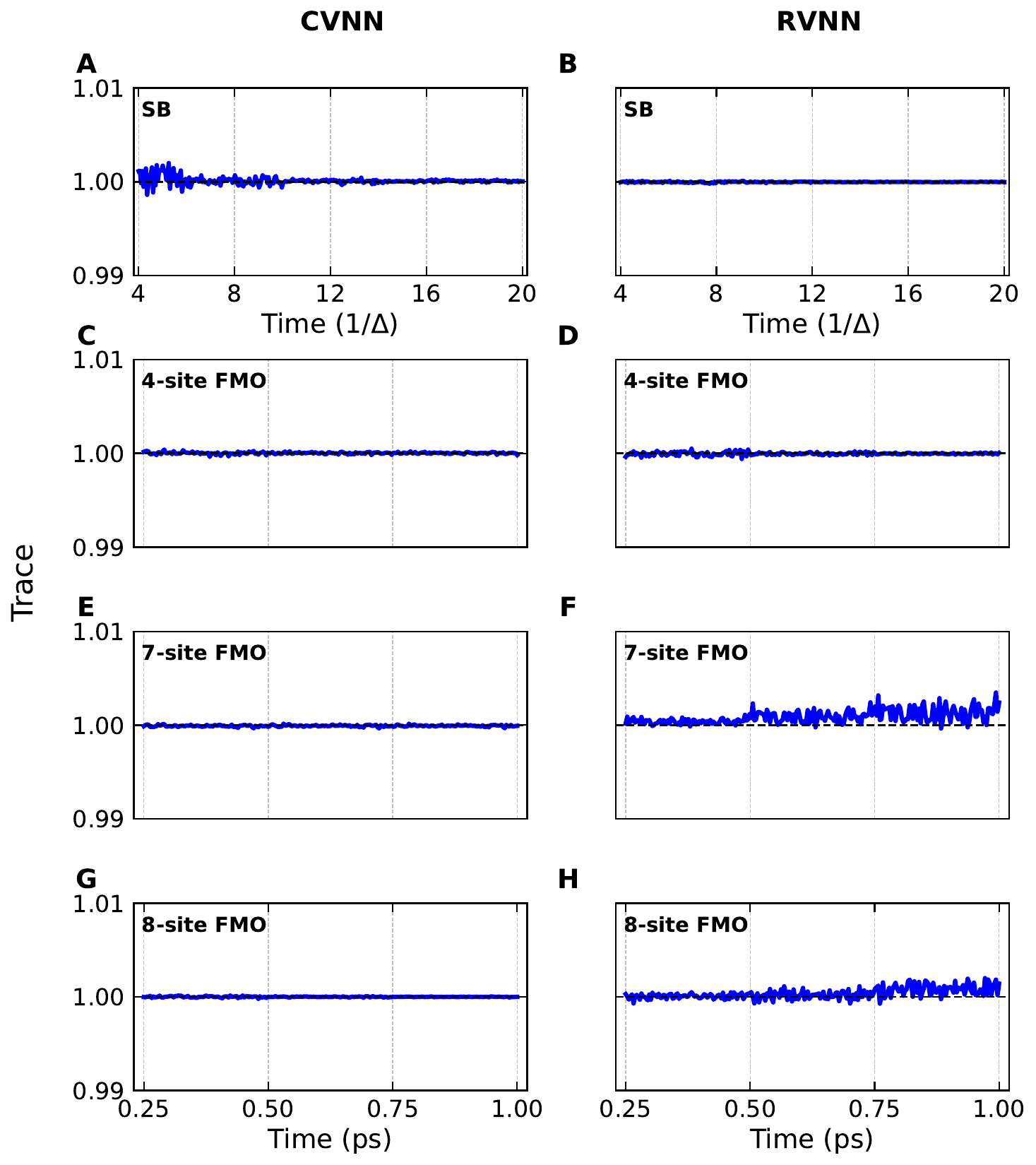}
\caption{\red{Comparison of trace conservation in predicted dissipative dynamics. Results for the CVNN (left) and RVNN (right) are shown for, from top to bottom, the SB model and the 4-site, 7-site, and 8-site FMO complexes (site-1). All predictions correspond to trajectories not seen during training. Parameters are: SB model—$\varepsilon/\Delta=0.0$, $\gamma/\Delta=9.0$, $\lambda/\Delta=6.0$, $\beta\Delta=1.0$; FMO complexes—4-site ($\gamma=250~\mathrm{cm}^{-1}$, $\lambda=70~\mathrm{cm}^{-1}$, $T=130~\mathrm{K}$), 7-site ($\gamma=350~\mathrm{cm}^{-1}$, $\lambda=70~\mathrm{cm}^{-1}$, $T=30~\mathrm{K}$), and 8-site ($\gamma=400~\mathrm{cm}^{-1}$, $\lambda=250~\mathrm{cm}^{-1}$, $T=30~\mathrm{K}$).}}
\label{fig:trace}
\end{figure}

%
%
\begin{table}[]
\centering
\small
\caption{\red{MAE in trace conservation, averaged over 100 trajectories, for RDMs predicted by CVNN and RVNN. The MAE measures the average deviation of the predicted trace from its theoretical value of 1.}}
\label{tab:mae_trace}
\begin{tabular}{|l|c|c|}
\hline
\textbf{Model} & \textbf{CVNN (MAE)} & \textbf{RVNN (MAE)} \\
\hline
SB Model & 2.92e-05 & 3.15e-05 \\
4-site FMO complex & 1.06e-04 & 2.09e-04 \\
7-site FMO complex & 5.48e-05 & 8.73e-05 \\
8-site FMO complex & 2.23e-05 & 8.52e-05 \\
\hline
\end{tabular}
\end{table}

For the SB model, each trajectory spans a total duration of $20/\Delta$ with a time step of $dt = 0.05$. In contrast, the FMO dynamics is restricted to a total time of $2~\mathrm{ps}$ with a time step of $0.005~\mathrm{ps}$. During training, both RVNN and CVNN models operate in a sequence-to-sequence forecasting mode, wherein a fixed-length history of RDMs is provided as input and multiple future time steps are predicted simultaneously in a single forward pass (“one-shot” prediction). Specifically, for the SB model, each training sample consists of 81 input time steps, and the network is trained to predict the subsequent 40 time steps. For the 4-site, 7-site, and 8-site FMO complexes, each training sample likewise contains 81 input time steps, with the models trained to predict the next 80 time steps.

All RVNN and CVNN models are trained for $10^{5}$ epochs using the Adam optimizer with a learning rate of $10^{-3}$. This unified training protocol enables a controlled and systematic comparison of the predictive accuracy and physical consistency of real- and complex-valued neural network architectures across quantum systems of increasing complexity.

%
\begin{figure}[h]
\includegraphics[width=0.45\textwidth]{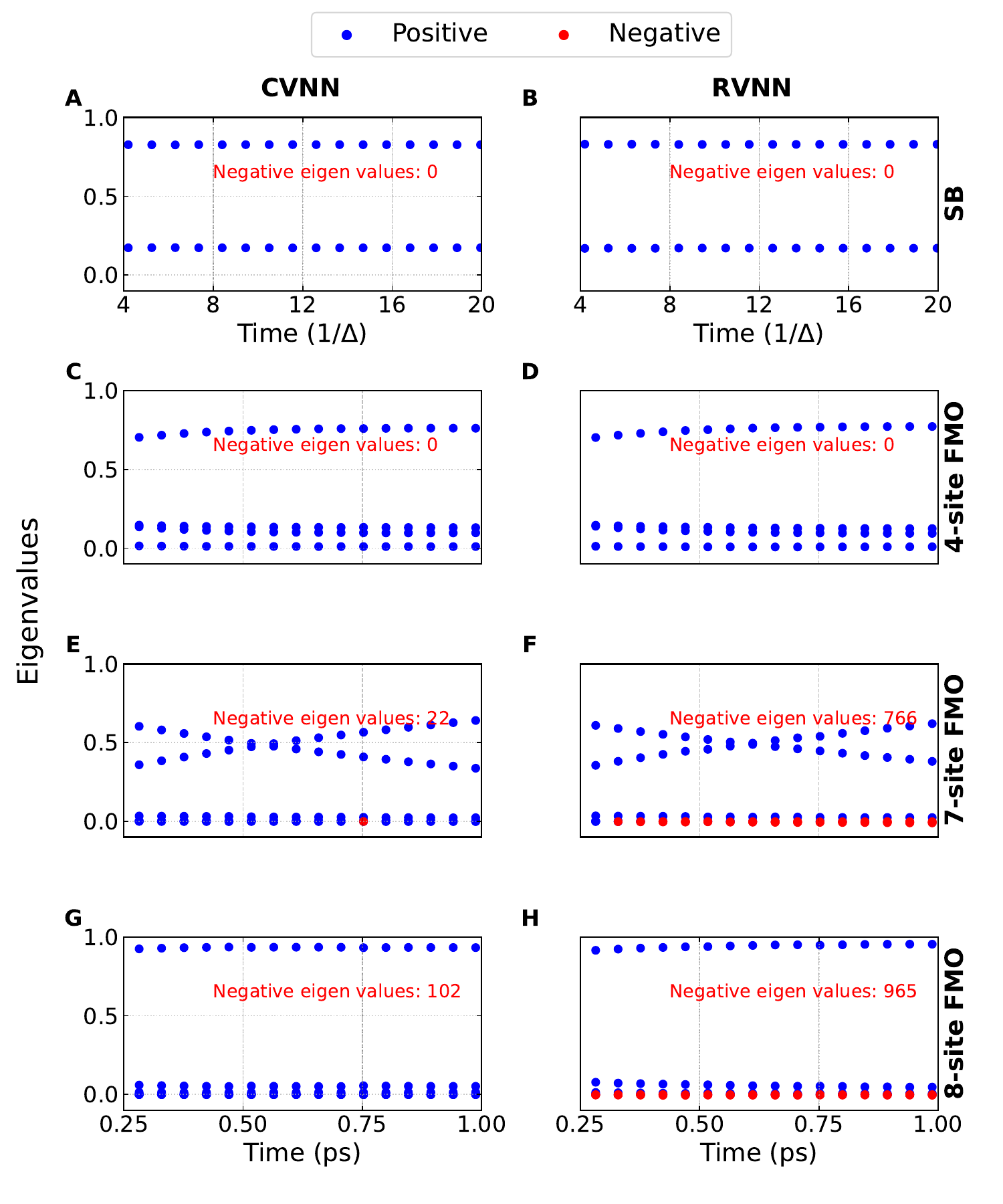}
\caption{\red{Evaluation of positive semi-definiteness of predicted RDMs via eigenvalue spectra for CVNN and RVNN models across all considered systems. Sparse positive eigenvalues are shown as blue dots for clarity, while negative eigenvalues are indicated in red, with their counts labeled for each model and system. Simulation parameters are consistent with those in Fig.~\ref{fig:trace}.}}
\label{fig:eigenvalues}
\end{figure}

As an initial step in comparing the two architectures, we assess the learning performance of the CVNN and RVNN by examining their training and validation loss evolution across the considered quantum systems (Fig.~\ref{fig:loss}). Analysis of the first 100 training epochs—a regime where key differences in learning efficiency emerge—reveals that both models exhibit a rapid initial loss decrease. For the low-dimensional SB model, where coherent effects are minimal, the CVNN and RVNN show nearly identical loss trajectories, reflecting the limited dynamical role of complex off-diagonal terms in the RDM.

This performance divergence intensifies with system size. For the 4- and 7-site FMO prototypes, the CVNN achieves a consistently faster reduction in both training and validation loss compared to the RVNN. This advantage is most pronounced for the largest, coherence-rich 8-site FMO complex, where the CVNN exhibits a substantially more rapid loss decrease. These results demonstrate that the CVNN's superior learning efficiency scales with system size, directly correlating with the increasing number and dynamical importance of complex-valued off-diagonal RDM elements.

We now turn to the physical fidelity of the predicted RDMs. We first assess trace conservation by comparing the CVNN and RVNN predictions. Fig.~\ref{fig:trace} shows the trace evolution for a representative test trajectory, while Table~\ref{tab:mae_trace} reports the mean absolute error (MAE) of the trace relative to unity, averaged over 100 trajectories.
\begin{table}[htbp]
\centering
\small
\caption{\red{Average negative eigenvalues calculated over 100 independent trajectories per model.}}
\label{tab:neg_eig}
\begin{tabular}{|l|c|c|}
\hline
\textbf{Model} & \textbf{CVNN} & \textbf{RVNN} \\
\hline
SB Model & 0.0 & 0.0 \\
4-site FMO complex & 38.05 & 40.63 \\
7-site FMO complex & 60.17 & 67.85 \\
8-site FMO complex & 53.87 & 73.71 \\
\hline
\end{tabular}
\end{table}

%
%
\begin{figure}
\begin{centering}
\includegraphics[width=0.45\textwidth]{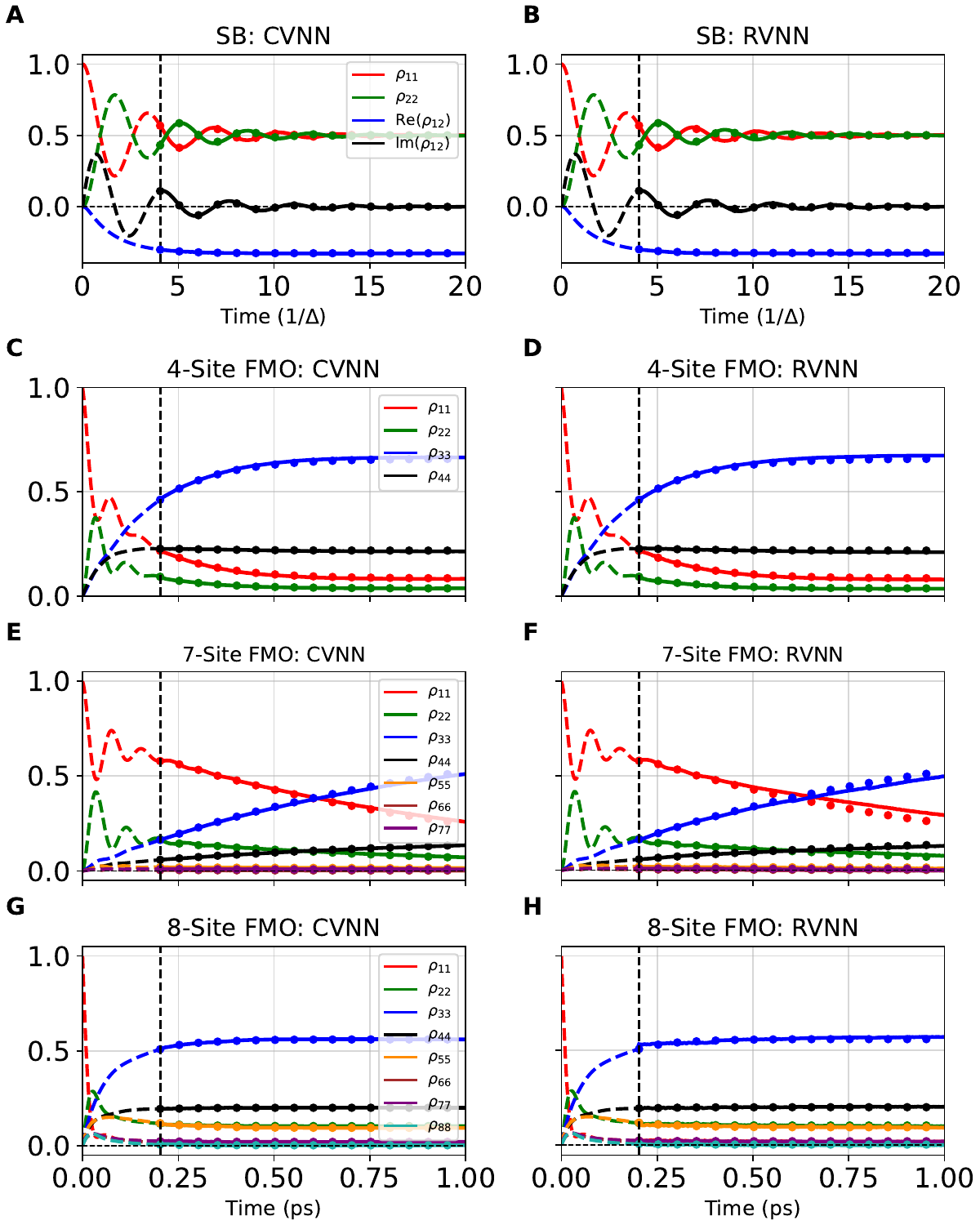}
\end{centering}
\caption{\red{Predicted time evolution of RDM elements. Panels (A, C, E, G) show CVNN results and panels (B, D, F, H) show RVNN results for the following systems: (A, B) SB model (populations and coherences); (C, D) populations of the 4-site FMO complex; (E, F) populations of the 7-site FMO complex; (G, H) populations of the 8-site FMO complex. Reference data are shown as dots. A vertical dashed line in each panel separates the provided input dynamics from the predicted dynamics. Coherence evolution for the FMO complexes is provided in the Supporting Information. Simulation parameters match those in Fig.~\ref{fig:trace}.}}
\label{fig:dynamics}
\end{figure}

%
\red{For the low-dimensional SB model, both CVNN and RVNN exhibit comparable trace MAEs on the order of $10^{-5}$. As shown in Table~\ref{tab:mae_trace}, the CVNN demonstrates a clear advantage for the FMO complexes, which becomes more systematic as the system representation becomes more complete. The 4-site FMO complex (comprising the four prominent sites of the full 7-site system) already shows a twofold trace accuracy improvement for the CVNN over the RVNN. For the complete 7-site complex, the CVNN maintains a 1.6x lower error. Most notably, for the 8-site complex, the CVNN's error reduces to $2.23 \times 10^{-5}$, which is nearly 4 times smaller than the RVNN's error ($ 8.52 \times 10^{-5}$).}

\red{We next evaluate positive semi-definiteness by analyzing the eigenvalue spectra of the predicted RDMs. Representative single-trajectory results are shown in Fig.~\ref{fig:eigenvalues}, and ensemble-averaged statistics over 100 trajectories are summarized in Table~\ref{tab:neg_eig}.

For the SB model, both CVNN and RVNN preserve positive semi-definiteness, with zero negative eigenvalues in the ensemble average. In contrast, violations emerge for the FMO complexes. For the 4-site FMO complex, both models exhibit a similar scale of violations, with the CVNN showing a reduction of approximately 6\% relative to the RVNN baseline (38.05 vs. 40.63). The advantage of the CVNN becomes more pronounced with increasing system complexity. For the 7-site FMO complex, the CVNN produces ~11\% fewer negative eigenvalues than the RVNN (60.17 vs. 67.85). This performance gap widens substantially for the 8-site complex, where the CVNN achieves a reduction of approximately 27\% relative to the RVNN (53.87 vs. 73.71).}

\begin{table}[]
\footnotesize 
\centering
\caption{\red{MAE for predicted RDM elements. For each model and system, the MAE (averaged over 100 trajectories and across all time steps) is reported separately for the diagonal elements and for the off-diagonal elements (calculated as the mean of the MAE for the real and imaginary components).}}
\label{tab:dyn_mae_compare}
\begin{tabular}{|l|cc|cc|}
\hline
\textbf{Model} 
& \multicolumn{2}{c|}{\textbf{CVNN}} 
& \multicolumn{2}{c|}{\textbf{RVNN}} \\
\cline{2-5}
& Diag & Off-diag & Diag & Off-diag \\
&  & (Real, Imag) &  & (Real, Imag) \\
\hline
SB model     & 7.01e-4 & (2.28e-3, 6.51e-4) & 7.85e-4 & (1.17e-3, 6.55e-4) \\
4-site FMO  & 1.94e-3 & (1.17e-3, 7.62e-5) & 2.87e-3 & (1.78e-3, 9.44e-5) \\
7-site FMO   & 5.23e-4 & (4.64e-4, 6.55e-5) & 1.23e-3 & (1.07e-3, 8.73e-5) \\
8-site FMO  & 4.87e-4 & (6.53e-4, 6.05e-5) & 1.40e-3 & (1.68e-3, 9.36e-5) \\
\hline
\end{tabular}
\end{table}

Taken together, the trace and eigenvalue analyses demonstrate that CVNNs preserve fundamental physical constraints at least as accurately as RVNNs, and in several cases significantly better. This improved physical consistency correlates with the enhanced learning efficiency observed in Fig.~\ref{fig:loss}, highlighting the advantage of operating natively in the complex domain for modeling larger and more coherent open quantum systems.

To further evaluate the CVNN and RVNN models, we compare their predictive accuracy for RDM dynamics. Fig.~\ref{fig:dynamics} shows the time evolution of diagonal (population) and off-diagonal (coherence) elements for a representative test trajectory. For clarity, both diagonal and off-diagonal elements are shown for the SB model, while for the FMO complexes the off-diagonal dynamics are provided in the Supporting Information (Section~S3). Ensemble-averaged accuracy is quantified in Table~\ref{tab:dyn_mae_compare}, which reports the MAE averaged over 100 independent test trajectories and all time steps, separately for diagonal and off-diagonal RDM elements. \red{A time resolved MAE trend is shown in Figs.~S2 and S3 (Section~S4 of the Supporting Information)}

\red{For the SB model, predictive accuracy is comparable between architectures. The CVNN reduces diagonal MAE by approximately 11\% (relative to the RVNN baseline), while the RVNN yields a ~49\% lower MAE for the real coherence components (relative to the CVNN).  These offsetting advantages indicate that neither model holds a decisive edge for this minimally coherent system. However, as system complexity increases, a clear CVNN advantage emerges. For the 4‑site FMO complex, the CVNN shows a consistent reduction in MAE: approximately 32\% lower for diagonal elements and 34\% lower for the real parts of coherences compared to the RVNN (relative to the RVNN baseline).

This advantage becomes more pronounced for the larger systems. In the 7‑site FMO complex, the CVNN’s diagonal error is less than half that of the RVNN (5.23e‑4 vs. 1.23e‑3), and its error for the real part of coherences is ~57\% lower (4.64e‑4 vs. 1.07e‑3). The trend continues for the 8‑site FMO complex, where the CVNN again yields substantially lower errors: roughly 65\% lower for diagonals and 61\% lower for the real coherence components. Together, these results demonstrate that the CVNN’s accuracy advantage—especially in capturing populations and key coherences—grows systematically as the open quantum system increases in size and coherence.}

\red{To demonstrate the general applicability and superior performance of the CVNN framework, we extend our analysis to the 8-site FMO complex modeled with an Ohmic spectral density. The results, presented in Supplementary Information (Section~S5), corroborate our primary findings and confirm the consistent advantage of CVNNs over RVNNs across different environmental models.

We further verified the robustness of the CVNN's superior performance by testing three different random seeds (42, 100, and 123). As detailed in Section~S6 of the Supporting Information, the CVNN outperformed the RVNN across all metrics—training/validation loss, trace conservation, average negative eigenvalues, and RDM prediction accuracy—for every seed tested. This seed-invariance analysis confirms that the CVNN's enhanced accuracy and physical consistency are not artifacts of initialization but are intrinsic to the complex-valued architecture.}

Overall, all these results demonstrate that CVNNs not only match or exceed RVNNs in predictive accuracy but also offer improved robustness with respect to fundamental quantum constraints, particularly as system size and coherence complexity increase.

\section{Concluding Remarks}

This work demonstrates that aligning ML architectures with the inherent mathematical structure of the physical systems they model is critical for performance. Quantum dissipative dynamics is fundamentally complex-valued, with essential information encoded in coupled amplitude-phase relationships and quantum coherences. Our results establish that neural networks operating natively in the complex domain offer a more natural and effective framework for learning such dynamics than conventional real-valued architectures, leading to superior accuracy and physical consistency.

In principle, quantum neural networks would provide the most faithful representation of quantum evolution. However, current NISQ hardware remains too constrained to offer practical advantages for this task. This limitation underscores the need for classical models that can capture essential quantum features without relying on quantum hardware. In this context, CVNNs emerge as a compelling intermediate paradigm—bridging the gap between purely real-valued models and fully quantum approaches.

By directly preserving the algebraic structure of complex-valued quantum states, CVNNs enable richer and more physically constrained representations. This translates to enhanced learning efficiency, greater stability with increasing system size, and more accurate adherence to quantum constraints as coherence complexity grows. More broadly, our findings position complex-valued ML as a promising and scalable direction for quantum-aware classical computation. It provides a hardware-independent pathway for modeling open quantum dynamics and can serve as an informative precursor to future hybrid classical–quantum strategies.

%
%
\section{SUPPLEMENTARY MATERIAL}
The supplementary material, provided as a separate file, includes additional results and further details on model architectures, parameters range and the farthest-point sampling strategy, the dynamics of off-diagonal elements (corresponding to Fig.~\ref{fig:dynamics}), a time-resolved MAE analysis, a robustness test on 8-site FMO complex with Ohmic spectral density, and a seed-invariance study. 

\section{Acknowledgments}

A.U. acknowledges funding from the National Natural Science Foundation of China (No. W2433037) and the Natural Science Foundation of Anhui Province (No. 2408085QA002). M.Y. acknowledges funding from Scientific Research Projects of Anhui Provincial Department of Education under Grant Nos. 2025AHGXZK20126 and 2025AHGXZK50054.   

\section{Data availability}
The code and data supporting this work are available at \url{https://github.com/Arif-PhyChem/cvnn}.

\section{Competing interests}
The authors declare no competing interests.

\section*{References}
\bibliographystyle{vancouver}
\bibliography{main.bib,references.bib}

\end{document}


\title{Supporting Information \\ for \\ Toward Quantum-Aware Machine Learning: Improved Prediction of Quantum Dissipative Dynamics via Complex Valued Neural Networks}

\author{Muhammad Atif}
\affiliation{School of Physics, 
Anhui University, Hefei 230601, China}

\author{Arif Ullah}
\email{arif@ahu.edu.cn}
\affiliation{School of Physics, 
Anhui University, Hefei 230601, China}

\author{Ming Yang}
\email{mingyang@ahu.edu.cn}
\affiliation{School of Physics, 
Anhui University, Hefei 230601, China}

\date{\today}

\maketitle

\section{Model Specifications and Computational Costs}
The architectural configurations and training hyperparameters for all models are detailed in the tables below. Table~\ref{tab:params_sb_4site} summarizes the specifications for the SB model and the 4‑site FMO complex, while Table~\ref{tab:params_7_8sites} presents the analogous details for the 7‑site and 8‑site FMO complexes. Both tables provide a side‑by‑side comparison of the CVNN and RVNN architectures, including network dimensions, total parameter counts, fixed training settings (epochs, Adam optimizer, learning rate), and the measured training and inference times.

\begin{table}[htb]\label{tab:params_sb_4site}
\small 
\centering
\begin{tabular}{|l|c|c|c|c|}
\hline
\textbf{Model}                & \textbf{CVNN (SB)}             & \textbf{RVNN (SB)}             & \textbf{CVNN (4-site)}     & \textbf{RVNN (4-site)}    \\ \hline
\multicolumn{5}{|c|}{\textbf{Training Hyperparameters}} \\ \hline
\textbf{Number of Epochs}         & 100000                           & 100000                           & 100000                           & 100000                           \\ \hline
\textbf{Learning Rate}            & 0.001                          & 0.001                           & 0.001                          & 0.001                           \\ \hline
\textbf{Optimizer}                & Adam                           & Adam                            & Adam                            & Adam                            \\ \hline

\multicolumn{5}{|c|}{\textbf{Model Specifications}} \\ \hline
\textbf{Number of Layers}         & 5                               & 5                                & 5                               & 5                                \\ \hline
\textbf{Input Size}               & 243                             & 324                             & 810                             & 1296                             \\ \hline
\textbf{Hidden Layers Size}        & 42                              & 64                               & 64                              & 82                               \\ \hline
\textbf{Output Size}              & 120                             & 160                             & 800                             & 1280                             \\ \hline
\textbf{Activation Function}      & CReLU                           & ReLU                            & CReLU                           & ReLU                            \\ \hline
\textbf{Number of Parameters}     & 41652                         & 43680                         & 232768                  & 233012                   \\ \hline

\multicolumn{5}{|c|}{\textbf{Training Costs}} \\ \hline
\textbf{Total Training Time (s)}  & 895.45                  &  438.41                         & 2517.15                   & 1014.76                   \\ \hline
\textbf{Time per Epoch (s)}       & 0.01                   & 0.004                        & 0.025                   & 0.01                   \\ \hline

\multicolumn{5}{|c|}{\textbf{Inference Costs}} \\ \hline
\textbf{Total Inference Time (s) (100 Trajectories)} & 0.25                 & 0.05                          & 0.19                  & 0.18                   \\ \hline
\textbf{Time per Trajectory (s)}   & 2.5e-3                   & 5.2e-4                          & 1.9e-3                  & 1.8e-3                 \\ \hline

\end{tabular}
\caption{Model configurations for the SB model and 4-site FMO complex. A side-by-side comparison of CVNN and RVNN architectures showing network dimensions, parameter counts, fixed training settings (epochs, Adam, learning rate), and measured training/inference times.}
\end{table}

\begin{table}[htb]\label{tab:params_7_8sites}
\small 
\centering
\begin{tabular}{|l|c|c|c|c|}
\hline
\textbf{Model}                & \textbf{CVNN (7-site)}             & \textbf{RVNN (7-site)}             & \textbf{CVNN (8-site)}     & \textbf{RVNN (8-site)}    \\ \hline
\multicolumn{5}{|c|}{\textbf{Training Hyperparameters}} \\ \hline
\textbf{Number of Epochs}         & 100000                           & 100000                           & 100000                           & 100000                           \\ \hline
\textbf{Learning Rate}            & 0.001                          & 0.001                           & 0.001                          & 0.001                           \\ \hline
\textbf{Optimizer}                & Adam                           & Adam                            & Adam                            & Adam                            \\ \hline

\multicolumn{5}{|c|}{\textbf{Model Specifications}} \\ \hline
\textbf{Number of Layers}         & 5                               & 5                                & 5                               & 5                                \\ \hline
\textbf{Input Size}               & 2268                             & 3969                             & 2916                             & 5184                             \\ \hline
\textbf{Hidden Layers Size}        & 108                              & 128                               & 111                              & 128                               \\ \hline
\textbf{Output Size}              & 2240                             & 3920                             & 2880                             & 5120                             \\ \hline
\textbf{Activation Function}      & CReLU                           & ReLU                            & CReLU                           & ReLU                            \\ \hline
\textbf{Number of Parameters}     & 1049056                         & 1063376                         & 1367276                  & 1373696                   \\ \hline

\multicolumn{5}{|c|}{\textbf{Training Costs}} \\ \hline
\textbf{Total Training Time (s)}  & 6482.21                  &  4496.09                         & 7897.77                   & 6712.18                   \\ \hline
\textbf{Time per Epoch (s)}       & 0.06                   & 0.04                        & 0.08                   & 0.07                   \\ \hline

\multicolumn{5}{|c|}{\textbf{Inference Costs}} \\ \hline
\textbf{Total Inference Time (s) (100 Trajectories)} & 0.28                  &  0.13                         & 0.30                  & 0.18                   \\ \hline
\textbf{Time per Trajectory (s)}   & 2.8e-3                   & 1.3e-3                          & 3.0e-3                   & 1.8e-3                  \\ \hline

\end{tabular}
\caption{Model configurations for the 7-site and 8-site FMO complexes. A side-by-side comparison of CVNN and RVNN architectures showing network dimensions, parameter counts, fixed training settings (epochs, Adam, learning rate), and measured training/inference times.}
\end{table}

\section{Training Data and Sampling Protocol}

We employ the SB dataset from our recently published QD3SET-1 database~\cite{ullah2023qd3set}. This dataset comprises 1000 stochastic trajectories generated for each combination of the dimensionless parameters 
\begin{equation}
    \tilde{\epsilon} = \epsilon/\Delta \in \{0, 1\}, \quad
\tilde{\lambda} = \lambda/\Delta \in \{0.1, 0.2, \ldots, 1.0\}, \quad
\tilde{\gamma} = \gamma/\Delta \in \{1, 2, \ldots, 10\},
\end{equation}
and
\begin{equation}
\tilde{\beta} = \beta \Delta \in \{0.1, 0.25, 0.5, 0.75, 1\},
\end{equation}
where $\epsilon$ is the bias, $\lambda$ is the reorganization energy, $\gamma$ is the bath relaxation rate, $\beta$ is the inverse temperature, and the tunneling matrix element $\Delta$ is used as the unit of energy. Quantum dynamics are generated using the hierarchical equations of motion (HEOM) method as implemented in the \textsc{QuTiP} software package~\cite{johansson2012qutip}. Each trajectory is propagated up to a final time $t\Delta = 20$ with a time step of $dt\Delta = 0.05$.

For the FMO complexes, excitonic dynamics are generated over a representative subset of the parameter space defined by the bath reorganization energy $\lambda$, bath relaxation rate $\gamma$, and temperature $T$. Specifically, the 500 most distant parameter combinations are selected from the sets
\begin{equation}
    \lambda \in \{10, 40, 70, \ldots, 520\}\,\mathrm{cm}^{-1}, \quad
\gamma \in \{25, 50, 75, \ldots, 500\}\,\mathrm{cm}^{-1}, \quad
T \in \{30, 50, 70, \ldots, 510\}\,\mathrm{K}.
\end{equation}
For each selected parameter combination, trajectories are generated for all considered initial excitation conditions (sites 1, 6, and 8). The quantum dynamics are propagated up to $t = 50$~ps with a time step of $dt = 5$~fs using the trace-conserving local thermalizing Lindblad master equation (LTLME) approach~\cite{mohseni2008environment, joseph2020thesis}, as implemented in the \textsc{quantum\_HEOM} package~\cite{joseph2019quant}.

To ensure representative and well-separated coverage of the bath parameter space while minimizing redundancy, we employ farthest point sampling (FPS)~\cite{dral2019mlatom, ullah2022predicting} to select trajectories for training and testing. This strategy improves generalization by constructing a training set that uniformly explores high-dimensional parameter regions, reducing bias toward densely sampled areas.

For the SB model, each trajectory corresponds to a point in a three-dimensional parameter space,
\begin{equation}
    \mathbf{x} = (\lambda, \gamma, \beta) \in \mathbb{R}^3.
\end{equation}
 Given a finite candidate set $\mathcal{P} = \{\mathbf{x}_i\}_{i=1}^{N}$, FPS iteratively constructs a subset $\mathcal{S} \subset \mathcal{P}$ by maximizing the minimum distance to the already selected points. Starting from an initial seed point $\mathbf{x}_1$, the $k$-th point is chosen according to
\begin{equation}
    \mathbf{x}_k = \arg\max_{\mathbf{x} \in \mathcal{P} \setminus \mathcal{S}_{k-1}}
    \Bigl(
        \min_{\mathbf{y} \in \mathcal{S}_{k-1}} \|\mathbf{x} - \mathbf{y}\|_2
    \Bigr),
\end{equation}
where $\mathcal{S}_{k-1} = \{ \mathbf{x}_1, \ldots, \mathbf{x}_{k-1} \}$ denotes the set of already selected points. In addition, $\mathbf{x}$ represents a candidate trajectory in the full dataset, while $\mathbf{y}$ represents a previously selected trajectory in the sampled subset. The Euclidean norm $\|\cdot\|_2$ measures the distance between two points in parameter space.

An analogous FPS strategy is employed for the FMO complexes, where each trajectory is represented by a point in the three-dimensional parameter space
\begin{equation}
    \mathbf{x} = (\lambda, \gamma, T),
\end{equation}
where the same iterative maximization of the minimum Euclidean distance is applied to select the most widely separated parameter combinations. 

By construction, FPS yields a training set that uniformly explores the high-dimensional bath parameter space, improving generalization and reducing bias toward densely populated regions. This is particularly important for learning quantum dissipative dynamics, where model performance can be highly sensitive to bath characteristics and temperature-dependent effects.

Based on the FPS selection, in our study, for each system—the SB model and the three FMO variants—400 trajectories are designated for training, while the remaining trajectories are held out exclusively for out-of-sample testing.

\section{Evolution of Key RDM Coherences}
\begin{figure}[h!]
\centering
\includegraphics[width=0.65\textwidth]{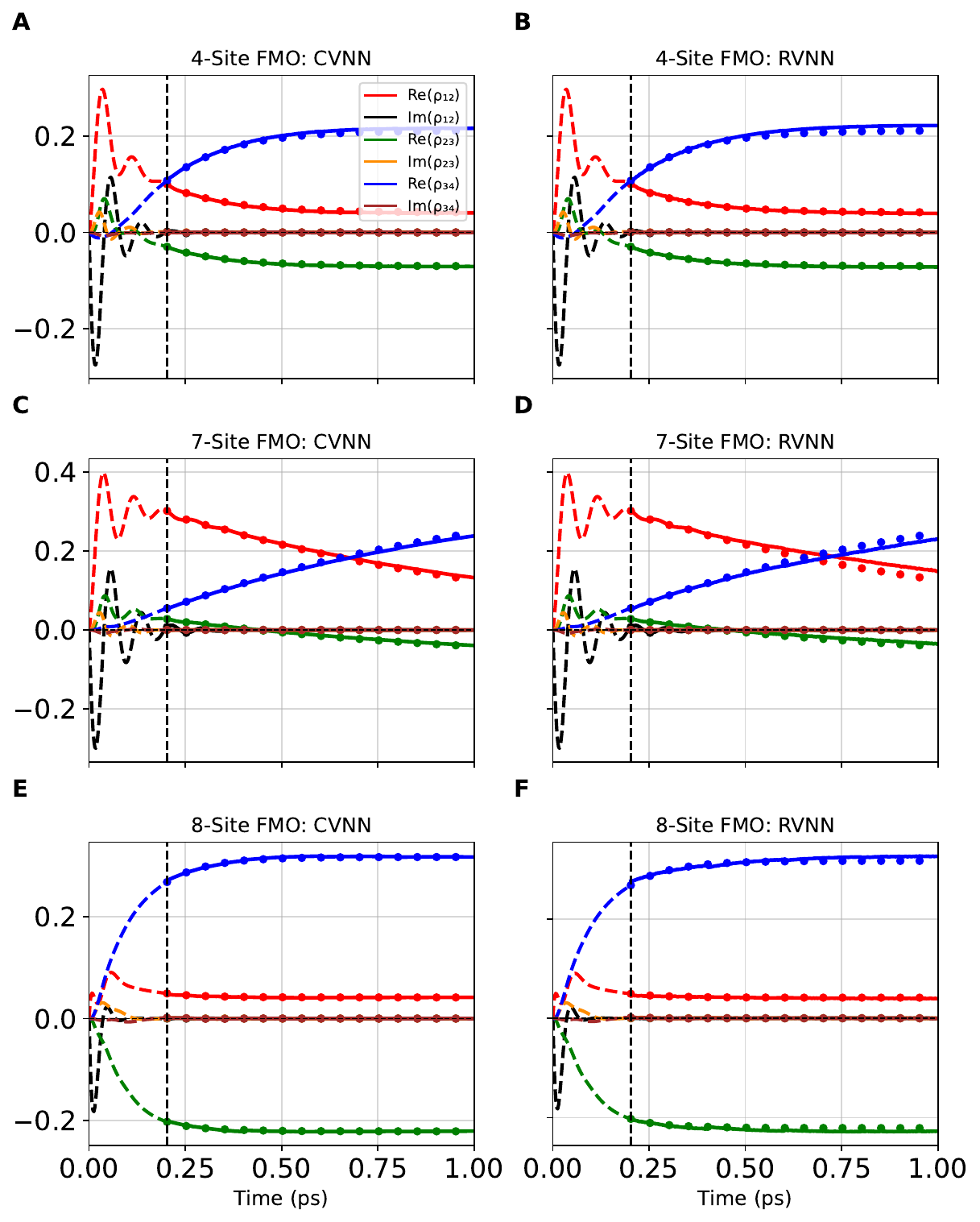}
\caption{Time evolution of selected prominent RDM off-diagonal (coherence) terms predicted by CVNN and RVNN models. Each row corresponds to an FMO complex: 4-site (A, B), 7-site (C, D), and 8-site (E, F). The left column (A, C, E) displays CVNN predictions; the right column (B, D, F) displays RVNN predictions. Reference dynamics are overlaid as dots. A vertical dashed line in each panel separates the provided input dynamics from the predicted dynamics. Parameters: 4-site ($\gamma=250\ \mathrm{cm}^{-1}$, $\lambda=70\ \mathrm{cm}^{-1}$, $T=130\ \mathrm{K}$); 7-site ($\gamma=350\ \mathrm{cm}^{-1}$, $\lambda=70\ \mathrm{cm}^{-1}$, $T=30\ \mathrm{K}$); 8-site ($\gamma=400\ \mathrm{cm}^{-1}$, $\lambda=250\ \mathrm{cm}^{-1}$, $T=30\ \mathrm{K}$).}
\label{fig:off-diag}
\end{figure}

The time evolution of selected, prominent off-diagonal RDM elements—representing quantum coherences—is shown in Fig.~\ref{fig:off-diag} for the CVNN and RVNN models, corresponding to the test trajectory presented in the main text. The evolution of the diagonal elements is provided therein.

\section{Time-Resolved MAEs}
\begin{figure}[htb]
\centering
\includegraphics[width=0.65\textwidth]{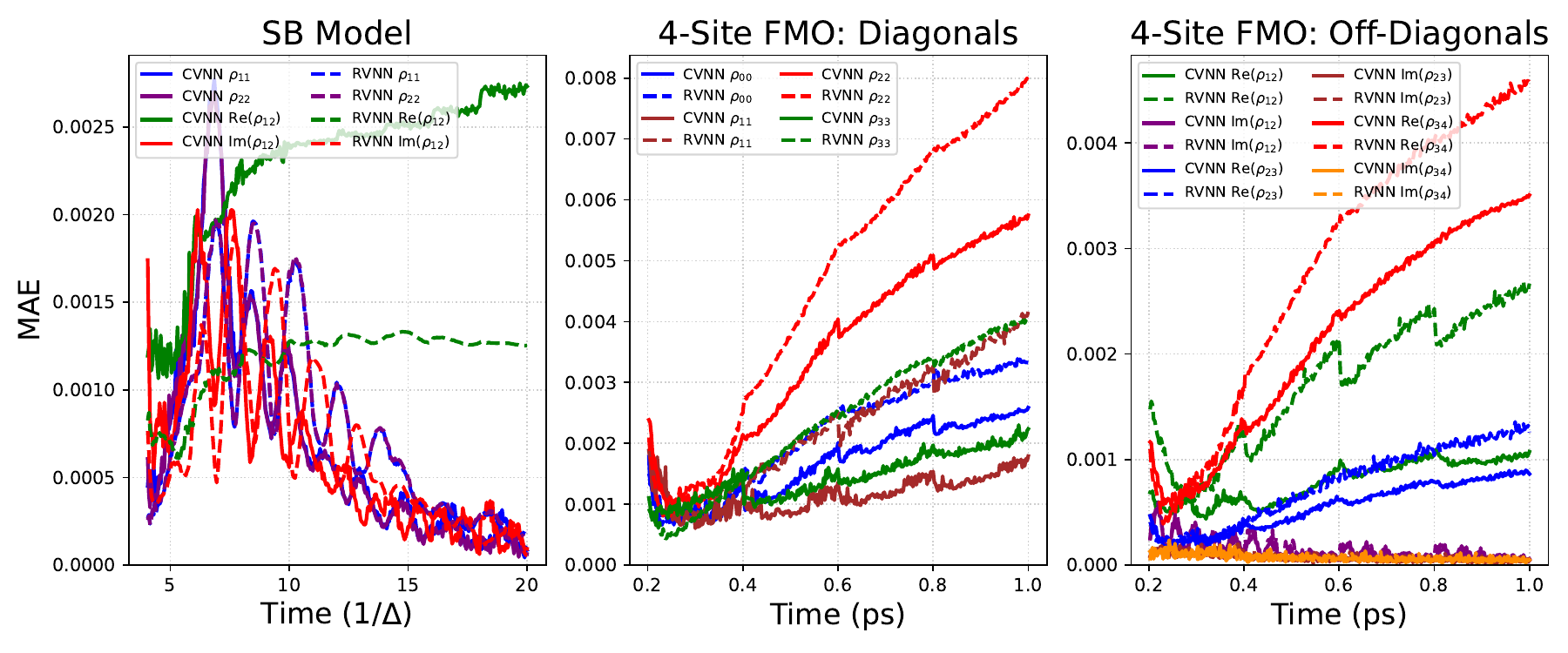}
\caption{Comparison of the time-resolved MAE for CVNN and RVNN across SB model and 4-site FMO complex.}
\label{fig:sb_4fmo_mae}
\end{figure}

\begin{figure}[htb]
\centering
\includegraphics[width=0.65\textwidth]{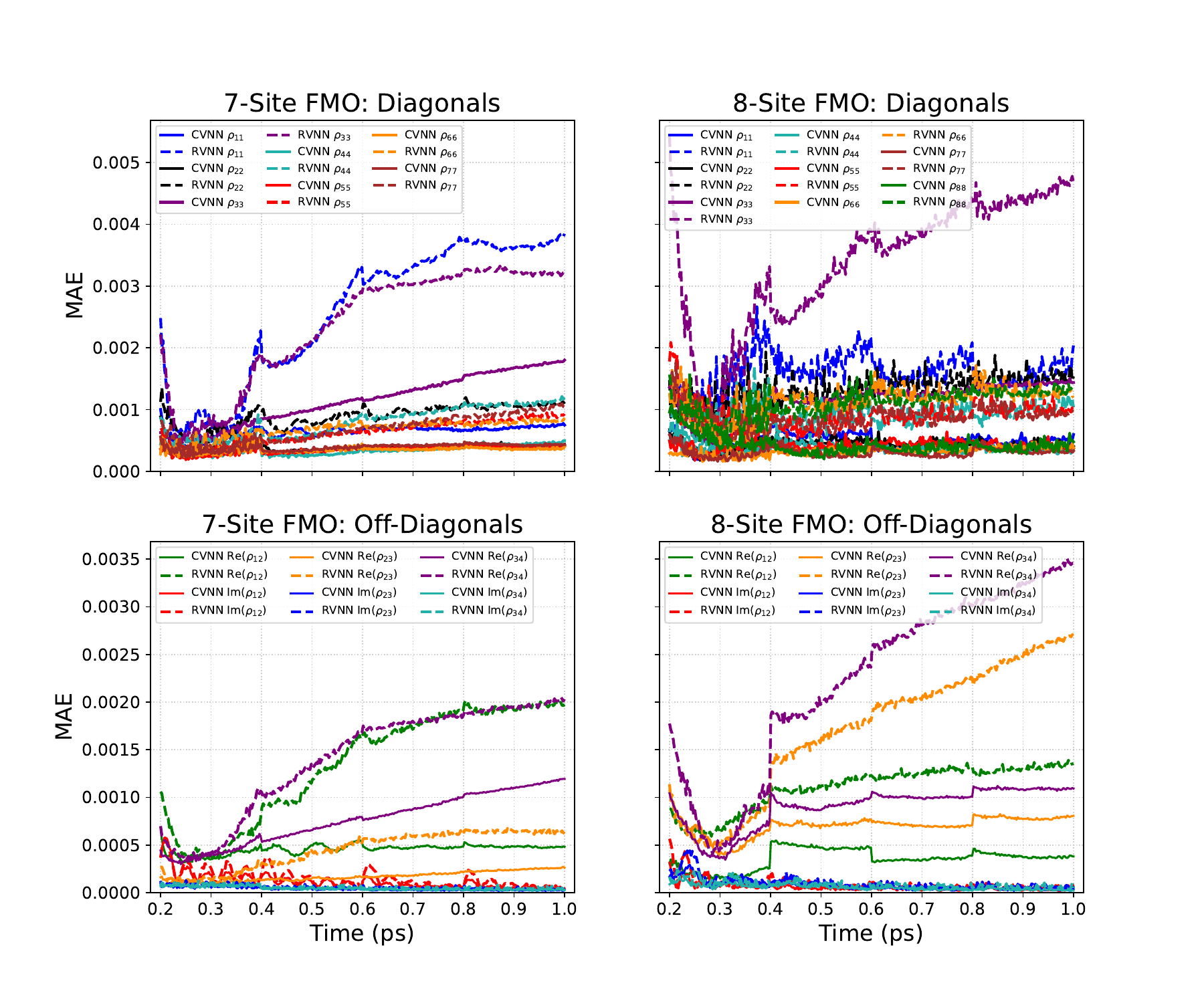}
\caption{Comparison of the time-resolved MAE for CVNN and RVNN across 7-site and 8-site FMO complexes.}
\label{fig:7_8_fmo_mae}
\end{figure}

This section presents a time-resolved evaluation of prediction accuracy based on the MAE of the RDM elements. Figs.~\ref{fig:sb_4fmo_mae} and \ref{fig:7_8_fmo_mae} depict the temporal evolution of the MAE for both CVNN and RVNN models across all studied systems, averaged over 100 test trajectories, thereby complementing the time-averaged metrics discussed in the main text. Overall, the CVNN exhibits consistently lower MAE throughout the trajectories for most systems and RDM components. The primary exception occurs in the low-coherence SB model, where the real part of the off-diagonal element shows comparable or slightly higher errors relative to the RVNN.

\section{Results for the 8-Site FMO Complex with Ohmic Spectral Density}

In this section, we present results for the 8-site FMO complex obtained using an Ohmic spectral density. The bath spectral density is defined as
\begin{equation}
J(\omega)= \pi \lambda \frac{\omega}{\gamma} e^{-\omega/\gamma},
\end{equation}
where $\lambda$ denotes the bath reorganization energy and $\omega_c$ is the cutoff frequency characterizing the bath relaxation rate. The performance of CVNNs and RVNNs is systematically evaluated using both numerical accuracy and physically motivated consistency metrics.

The dataset for the Ohmic spectral density is generated using the same simulation protocol, Hamiltonian, and numerical settings as those employed for the 8-site FMO complex with Debye spectral density, as described in the main text and Section~S2. In particular, the system Hamiltonians are parameterized following the 8-site extension introduced by Jia \emph{et al.}\cite{jia2015hybrid}, consistent with the QD3SET-1 database\cite{ullah2023qd3set} construction. The quantum dynamics are propagated using a trace-conserving local thermalizing Lindblad master equation (LTLME)\cite{mohseni2008environment}. The associated parameter space $(\lambda, \gamma, T)$ spans the bath reorganization energy, relaxation rate, and temperature, respectively, yielding a total of 500 stochastic quantum trajectories for an initial excitation localized on site~1.

Based on farthest point sampling in parameter space, 400 trajectories are selected for training, while the remaining 100 trajectories are reserved exclusively for testing. This split ensures broad coverage of the bath parameter space while providing a stringent evaluation of the models’ generalization performance.

\begin{figure}[htb]
\includegraphics[width=0.5\textwidth]{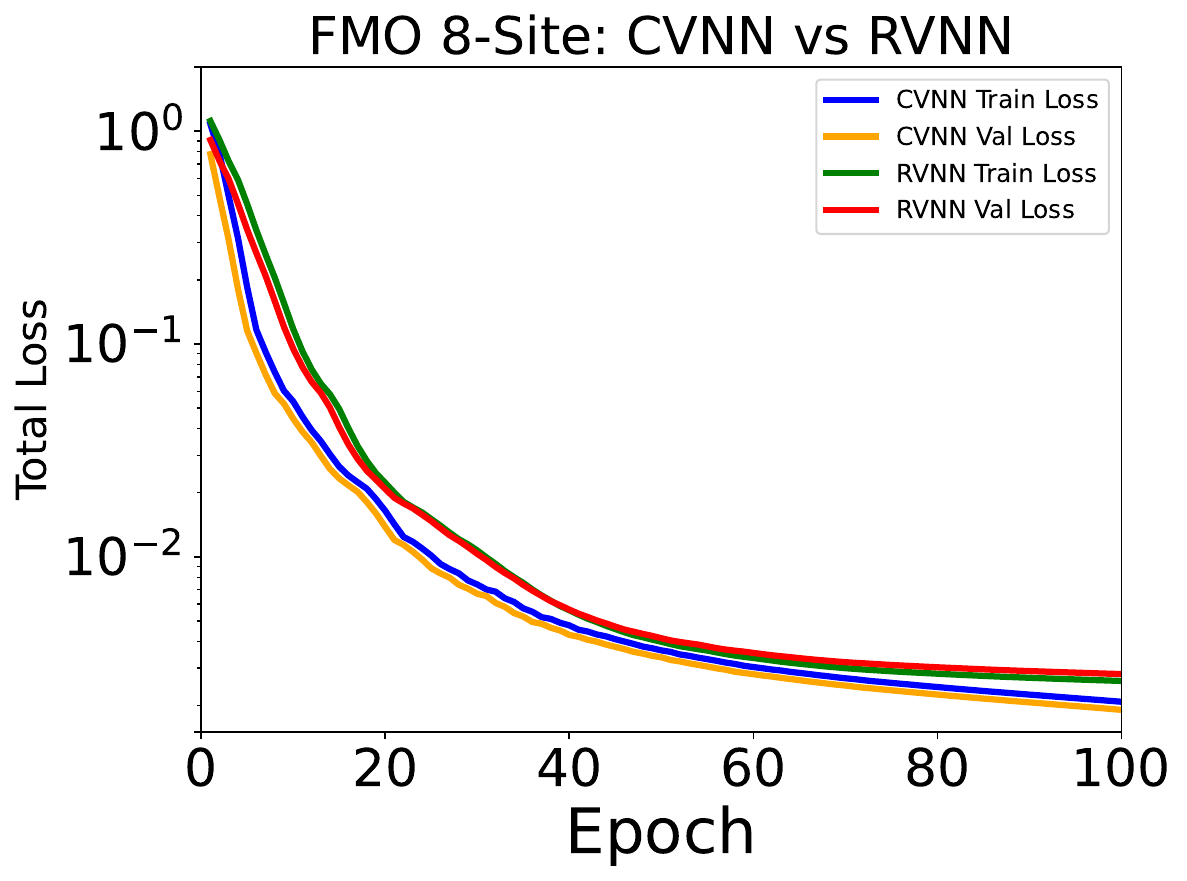}
\caption{A comparison of training and validation loss curves versus epoch number for the CVNN and RVNN models. Results are taken for 8-site FMO complex.}
\label{fig:ohmic_loss}
\end{figure}

Fig.~\ref{fig:ohmic_loss} shows the training and validation loss curves for both models. Similar to the Debye spectral density case discussed in the main text, the CVNN exhibits faster convergence and more efficient learning than the RVNN, indicating an improved representation of the underlying quantum dynamics.

The conservation of the trace and the preservation of positive semi-definiteness of the RDM are evaluated using the MAE in the trace and the average number of negative eigenvalues, respectively, computed over the 100 test trajectories. In addition, the accuracy of the predicted RDM dynamics is assessed through the MAE of both diagonal and off-diagonal matrix elements. These results are summarized in Table~\ref{tab:ohmic_8site_fmo_summary}.

Consistent with the Debye spectral density results, the CVNN systematically outperforms the RVNN across all evaluated metrics. In particular, the CVNN achieves improved trace conservation and exhibits fewer violations of positive semi-definiteness, reflecting enhanced physical fidelity. Moreover, the CVNN provides more accurate predictions for both population (diagonal) and coherence (off-diagonal) dynamics, highlighting the advantages of explicitly complex-valued architectures for modeling open quantum systems.

These findings demonstrate that the superior performance of CVNNs is robust with respect to the choice of spectral density and is not restricted to a specific environmental model, underscoring their broader applicability for learning quantum dissipative dynamics.

\begin{table}[htb]
\centering
\small
\caption{Quantitative comparison of CVNN and RVNN predictions for the 8-site FMO complex with Ohmic spectral density. Metrics are averaged over 100 independent trajectories and across all time steps. MAE values for RDM off-diagonal elements are reported separately for the real and imaginary components.}
\label{tab:ohmic_8site_fmo_summary}
\begin{tabular}{|l|c|c|c|c|c|c|}
\hline
\multirow{2}{*}{\textbf{Model}} & 
\multicolumn{2}{c|}{\textbf{Physical Constraints}} & 
\multicolumn{4}{c|}{\textbf{RDM Element Accuracy (MAE)}} \\
\cline{2-7}
& \textbf{Trace MAE} & \textbf{Avg. Negative Eigenvalues} & 
\textbf{Diagonal} & \multicolumn{2}{c|}{\textbf{Off-diagonal}} & \\
\cline{5-6}
& & & & \textbf{Real} & \textbf{Imaginary} & \\
\hline
CVNN & $3.80 \times 10^{-5}$ & 61.04 & $4.04 \times 10^{-4}$ & $4.98 \times 10^{-4}$ & $5.18 \times 10^{-5}$ & \\
\hline
RVNN & $9.93 \times 10^{-5}$ & 71.85 & $5.49 \times 10^{-4}$ & $7.01 \times 10^{-4}$ & $6.77 \times 10^{-5}$ & \\
\hline
\end{tabular}
\end{table}


\section{Evaluation of CVNN and RVNN Across Different Random Seeds}

To verify that the superior performance of CVNNs is not an artifact of a particular random seed, we performed an analysis using three different seeds: 42, 100, and 123. For each seed, we evaluated both CVNN and RVNN models on multiple metrics, including training and validation loss, trace conservation, average number of negative eigenvalues, and RDM prediction accuracy across 100 test trajectories.

\begin{figure}[htb]
\includegraphics[width=0.5\textwidth]{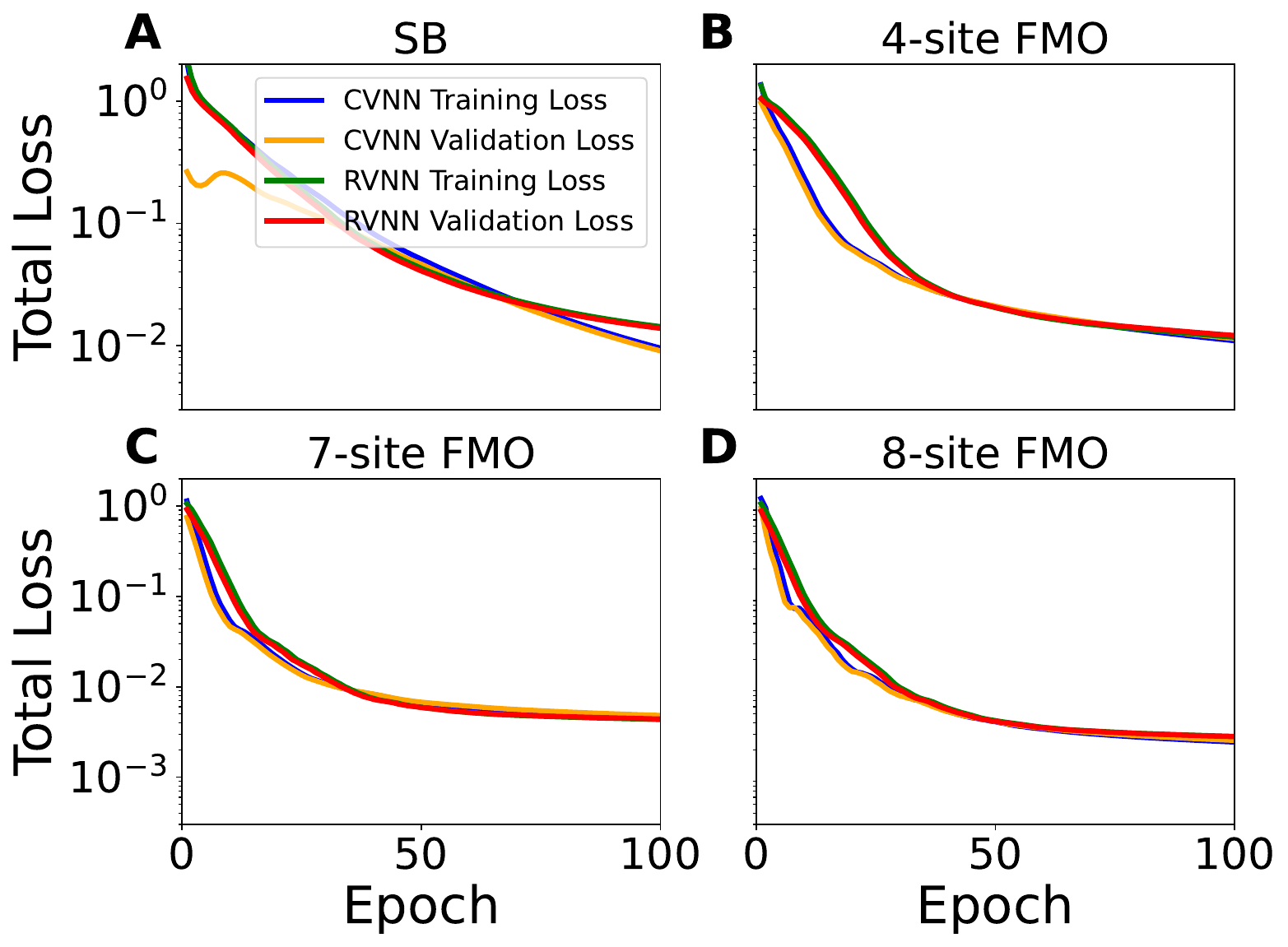}
\caption{A comparison of training and validation loss curves versus epoch number for the CVNN and RVNN models for random seed 42.}
\label{fig:loss_42}
\end{figure}

\begin{figure}[htb]
\includegraphics[width=0.5\textwidth]{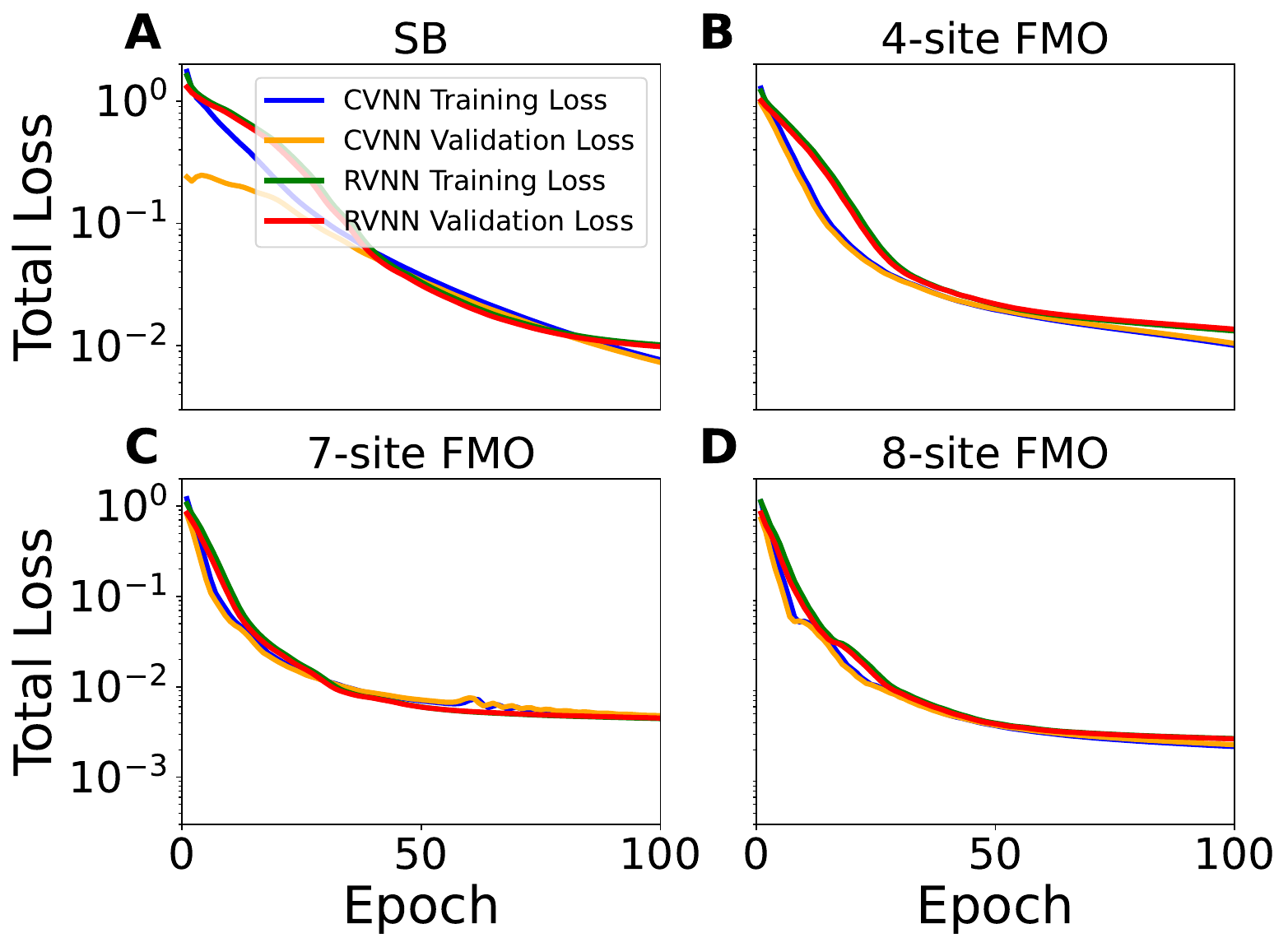}
\caption{A comparison of training and validation loss curves versus epoch number for the CVNN and RVNN models for random seed 100.}
\label{fig:loss_100}
\end{figure}

\begin{figure}[htb]
\includegraphics[width=0.5\textwidth]{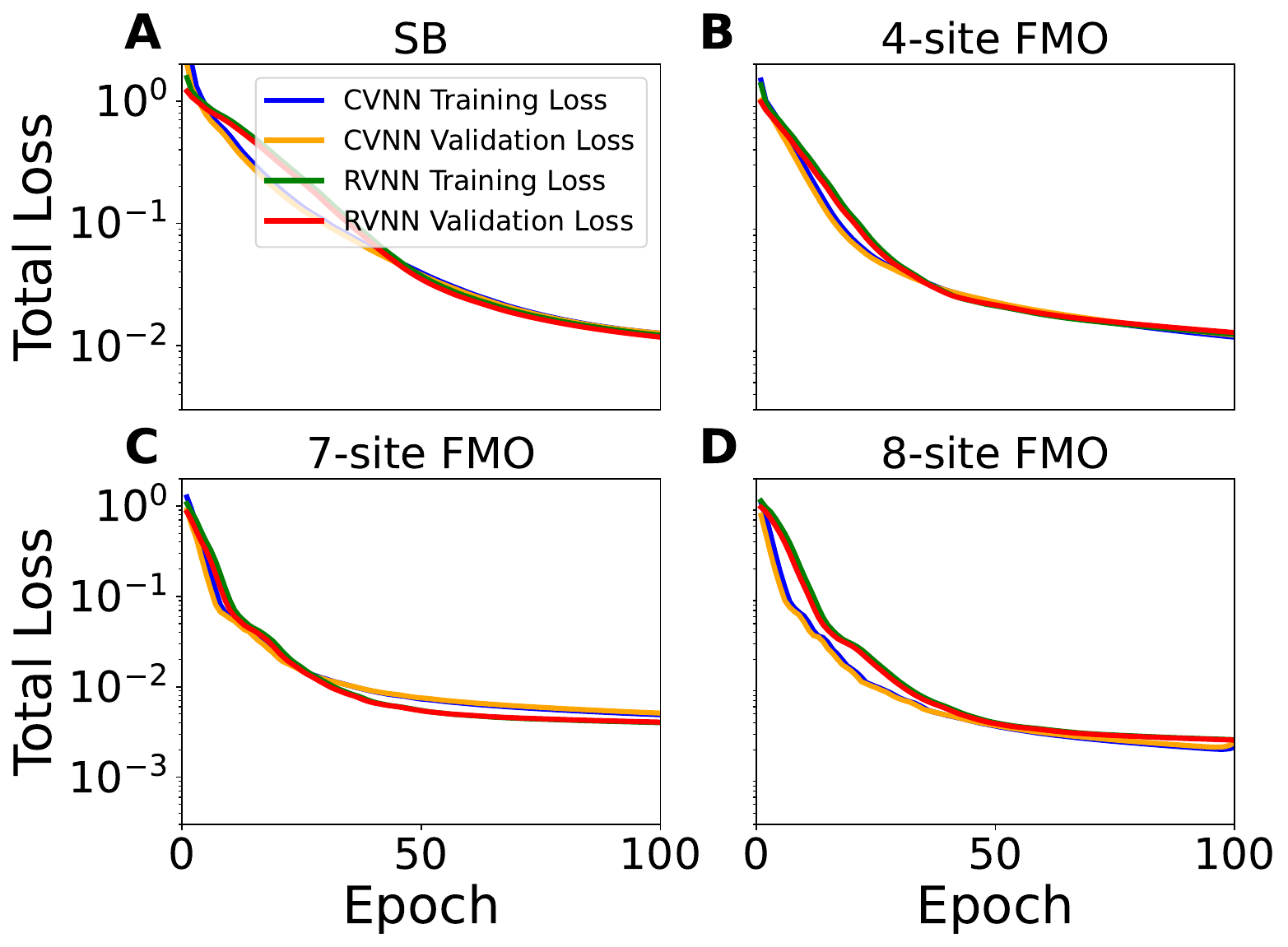}
\caption{A comparison of training and validation loss curves versus epoch number for the CVNN and RVNN models for random seed 123.}
\label{fig:loss_123}
\end{figure}

Figs.~\ref{fig:loss_42}, \ref{fig:loss_100}, and \ref{fig:loss_123} present the learning curves for all three seeds, illustrating consistent convergence behavior for both models. Trace conservation, quantified via MAE over the test trajectories, is summarized in Table~\ref{tab:seed_loss}, while the average count of negative eigenvalues is reported in Table~\ref{tab:seed_eigen}. Prediction accuracy for the RDM dynamics, measured as MAE over the 100 trajectories, is presented in Table~\ref{tab:seed_dyn}.

Across all random seeds, the CVNN consistently outperforms the RVNN, achieving better preservation of physical constraints and higher accuracy in predicting both population and coherence dynamics. These results demonstrate that the improved performance of CVNNs is robust and independent of the choice of random initialization.

\begin{table}[htb]\label{tab:seed_loss}
\centering
\small
\caption{MAE in trace conservation, averaged over 100 trajectories, for RDMs predicted by CVNN and RVNN. The MAE measures the average deviation of the predicted trace from its theoretical value of 1. Results are shown for different random seeds.}
\label{tab:mae_trace}
\begin{tabular}{|l|c|c|}
\hline
\textbf{Model} & \textbf{CVNN (MAE)} & \textbf{RVNN (MAE)} \\
\hline
\multicolumn{3}{|c|}{\textbf{Random Seed 42}} \\
\hline
SB Model           & 1.06e-04 & 1.20e-04 \\
4-site FMO complex & 7.72e-05 & 8.81e-05 \\
7-site FMO complex & 4.41e-05 & 6.64e-05 \\
8-site FMO complex & 6.98e-05 & 8.35e-05 \\
\hline
\multicolumn{3}{|c|}{\textbf{Random Seed 100}} \\
\hline
SB Model           & 5.86e-05 & 8.10e-05 \\
4-site FMO complex & 1.22e-05 & 7.93e-05 \\
7-site FMO complex & 6.66e-05 & 7.72e-05 \\
8-site FMO complex & 3.92e-05 & 4.60e-05 \\
\hline
\multicolumn{3}{|c|}{\textbf{Random Seed 123}} \\
\hline
SB Model           & 7.93e-05 & 1.19e-04 \\
4-site FMO complex & 1.11e-04 & 1.34e-04 \\
7-site FMO complex & 6.16e-05 & 7.19e-05 \\
8-site FMO complex & 5.00e-05 & 5.35e-05 \\
\hline
\end{tabular}
\end{table}

\begin{table}[htb]\label{tab:seed_eigen}
\centering
\small
\caption{Average negative eigenvalues calculated over 100 independent trajectories per model, shown for different random seeds.}
\label{tab:neg_eig}
\begin{tabular}{|l|c|c|}
\hline
\textbf{Model} & \textbf{CVNN} & \textbf{RVNN} \\
\hline
\multicolumn{3}{|c|}{\textbf{Random Seed 42}} \\
\hline
SB Model           & 0.00  & 0.00  \\
4-site FMO complex & 28.46 & 41.23 \\
7-site FMO complex & 35.41 & 76.68 \\
8-site FMO complex & 57.43 & 62.69 \\
\hline
\multicolumn{3}{|c|}{\textbf{Random Seed 100}} \\
\hline
SB Model           & 0.00  & 0.00  \\
4-site FMO complex & 29.09 & 38.71 \\
7-site FMO complex & 47.29 & 73.81 \\
8-site FMO complex & 42.18 & 50.75 \\
\hline
\multicolumn{3}{|c|}{\textbf{Random Seed 123}} \\
\hline
SB Model           & 0.00  & 0.00  \\
4-site FMO complex & 23.51 & 29.19 \\
7-site FMO complex & 69.37 & 76.83 \\
8-site FMO complex & 57.80 & 62.90 \\
\hline
\end{tabular}
\end{table}


\begin{table}[htb]\label{tab:seed_dyn}
\footnotesize
\centering
\caption{MAE for predicted RDM elements. For each model and system, the MAE (averaged over 100 trajectories and across all time steps) is reported separately for the diagonal elements and for the off-diagonal elements (calculated as the mean of the MAE for the real and imaginary components). Results are shown for different random seeds.}
\label{tab:dyn_mae_compare}
\begin{tabular}{|l|cc|cc|}
\hline
\textbf{Model} 
& \multicolumn{2}{c|}{\textbf{CVNN}} 
& \multicolumn{2}{c|}{\textbf{RVNN}} \\
\cline{2-5}
& Diag & Off-diag (Real, Imag) & Diag & Off-diag (Real, Imag) \\
\hline
\multicolumn{5}{|c|}{\textbf{Random Seed 42}} \\
\hline
SB model           & 1.34e-3 & (2.74e-3, 1.25e-4) & 1.76e-4 & (4.15e-3, 1.47e-4) \\
4-site FMO complex & 1.23e-3 & (7.24e-4, 7.46e-5) & 2.66e-3 & (1.74e-3, 7.68e-5) \\
7-site FMO complex & 6.52e-4 & (5.96e-4, 7.24e-5) & 9.91e-4 & (8.45e-4, 7.48e-5) \\
8-site FMO complex & 6.10e-4 & (8.51e-4, 9.50e-5) & 9.31e-4 & (1.37e-3, 1.12e-4) \\
\hline
\multicolumn{5}{|c|}{\textbf{Random Seed 100}} \\
\hline
SB model           & 8.83e-4 & (1.38e-3, 8.66e-4) & 9.83e-4 & (2.01e-3, 9.21e-4) \\
4-site FMO complex & 1.69e-3 & (1.01e-3, 9.38e-5) & 1.80e-3 & (1.16e-3, 9.95e-5) \\
7-site FMO complex & 6.33e-4 & (5.58e-4, 7.17e-5) & 1.31e-3 & (1.18e-3, 8.47e-5) \\
8-site FMO complex & 5.55e-4 & (6.44e-4, 6.72e-5) & 5.99e-4 & (8.65e-4, 6.95e-5) \\
\hline
\multicolumn{5}{|c|}{\textbf{Random Seed 123}} \\
\hline
SB model           & 1.70e-3 & (2.19e-3, 1.63e-3) & 2.27e-3 & (5.61e-3, 2.18e-3) \\
4-site FMO complex & 1.87e-3 & (1.19e-4, 1.16e-5) & 1.93e-3 & (1.24e-3, 8.34e-5) \\
7-site FMO complex & 6.63e-4 & (5.96e-4, 8.49e-5) & 1.20e-3 & (1.01e-3, 8.53e-5) \\
8-site FMO complex & 4.97e-4 & (6.09e-4, 5.94e-5) & 8.34e-4 & (9.69e-4, 9.22e-5) \\
\hline
\end{tabular}
\end{table}

\clearpage

\section*{References}
\bibliographystyle{vancouver}
\bibliography{main.bib,references.bib}
